\documentclass[a4paper,12pt,preprint,tightenlines,floatfix,showpacs,preprintnumbers,
nofootinbib,eqsecnum]{article}

\usepackage[utf8]{inputenc}
\usepackage{amssymb,amsmath,mathrsfs}
\usepackage[english]{babel}

\usepackage[a4paper, total={7in, 9in}]{geometry}
\usepackage{color}
\usepackage{epstopdf}
\usepackage{graphicx,wrapfig}
\usepackage{multirow}
\usepackage{mathtools}
\usepackage{tabularx}
\usepackage{makecell}
\usepackage{caption}
\usepackage{float}
\usepackage[colorlinks,citecolor=blue,urlcolor=blue,bookmarks=false
]{hyperref}
\RequirePackage{flushend}
\RequirePackage[numbers,sort&compress]{natbib}
\usepackage{verbatim}
\usepackage{authblk}
\usepackage[font=footnotesize,labelfont=bf,singlelinecheck=false]{caption}
\usepackage[binary-units]{siunitx}
\usepackage{tikz}
\usetikzlibrary{arrows,shadows}
\usepackage{setspace}
\usepackage{array}
\newcolumntype{x}[1]
{>{\raggedright}p{#1}}

\def\1{\hbox{{1}\kern-.25em\hbox{l}}}

\newcommand{\msbar}{\overline{\rm{MS}}}

\pdfoutput=1

\begin{document}
\pagenumbering{gobble}
\date{\today}

\title{\vspace{-1.5cm}\begin{flushright}
\small{INR-TH-2023-001}
\end{flushright} \vspace{1.5cm} \textbf{The generalized Crewther relation and  V-scheme: analytic $\mathcal{O}(\alpha^4_s)$ results in QCD and QED
}}

\author[1]{ A.~L.~Kataev\footnote{kataev@ms2.inr.ac.ru}}
\author[1, 2, 3]{ V.~S.~Molokoedov\footnote{viktor\_molokoedov@mail.ru}}

\affil[1]{{\footnotesize Institute for Nuclear Research of the Russian Academy of Science,
 117312, Moscow, Russia}}
\affil[2]{{\footnotesize Research Computing Center, Moscow State University, 119991, Moscow, Russia}}
\affil[3]{{\footnotesize Moscow Institute of Physics and Technology, 141700, Dolgoprudny, Moscow Region, Russia}}

\maketitle

\begin{abstract}

Using the analytical $\rm{\overline{MS}}$-scheme three-loop contribution to the perturbative Coulomb-like part of the static color potential 
of heavy quark-antiquark system, we obtain the analytical expression for the fourth-order $\beta$-function in the gauge-invariant effective V-scheme in the case of the generic simple gauge group. Also we present the Adler function of electron-positron annihilation into hadrons and the coefficient function of the Bjorken polarized sum rule in the  V-scheme up to $\alpha^4_s$ terms. We demonstrate that at this level of PT in this effective scheme
the $\beta$-function is factorized in the conformal symmetry breaking term of the generalized Crewther relation, which connects the flavor non-singlet contributions to the Adler and Bjorken polarized sum rule functions. We prove why this relation will be true in other gauge-invariant renormalization  schemes as well. 
The obtained results enable to reveal the difference 
between the V-scheme $\beta$-function in QED and the Gell-Man--Low $\Psi$-function. This distinction arises due to the presence of the light-by-light type scattering corrections first appearing in the static potential at the three-loop level.

\end{abstract}


\newpage

\section{Introduction}
\pagenumbering{arabic}
\pagestyle{plain}
\setcounter{page}{1}

As known the binding energy of quark-antiquark system in a color singlet state in QCD can be described by two terms, namely by the perturbative Coulomb-like contributions at short distances and substantially nonperturbative long-distance ones, modeling the confinement description.
Investigation of this phenomenon is actively underway by means of the lattice calculations, where the linear dependence on distance $r$ is predicted for nonperturbative part of the static potential (see e.g. \cite{Bali:2000gf, Karbstein:2018mzo, BornyakovVG:2023rci} and references therein). 
In its turn, the non–abelian analogue of the Coulomb potential of QED is determined in the framework of the perturbation theory (PT) and, for instance,  it is the main component upon studying spectroscopy of bound states like heavy quarkonia in the Non-Relativistic QCD \cite{Brambilla:1999qa, Brambilla:2004jw}. 

The static potential of interaction of the heavy quark-antiquark pair in general is defined via the vacuum expectation value of the gauge-invariant Wilson loop\footnote{More precisely through the limit of
the logarithm of the path-ordering Wilson
loop over a closed rectangular contour divided by the interaction time when this time tends to infinity.
}. The perturbative part of this potential is
now available in analytic form in the $\rm{\overline{MS}}$ renormalization scheme 
at the three-loop 
level. In the momentum representation its expression has the following form:
\begin{equation}
\label{V}
V(\vec{q}^{\; 2})=-\frac{4\pi C_F\alpha_s(\vec{q}^{\; 2})}{\vec{q}^{\; 2}}\bigg[1+a_1a_s(\vec{q}^{\; 2})+a_2a^2_s(\vec{q}^{\; 2})  
+ \bigg(a_3+\frac{\pi^2 C^3_A L}{8}\bigg)a^3_s(\vec{q}^{\; 2})+\mathcal{O}(a^4_s)\bigg],
\end{equation}
where $\alpha_s$ is the renormalized strong coupling constant in the  
$\rm{\overline{MS}}$-scheme, 
$a_s$=$\alpha_s/\pi$, $L$=$\log(\mu^2/\vec{q}^{\; 2})$, 
$\mu^2$ is the scale $\rm{\overline{MS}}$-scheme parameter of the dimensional regularization, $\vec{q}^{\;2}$ is the square of the Euclidean three-dimensional momentum. The limit when time $T\rightarrow\infty$ formally leads to $q_0\rightarrow 0$ and the square of the Euclidean four-dimensional transferred momentum $Q^2\rightarrow \vec{q}^{\;2}$. Thus, technically we carry out the transition from the Euclidean four-dimensional space to its three-dimensional subspace. The renormalization group (RG)
uncontrollable logarithmic term \cite{Brambilla:1999qa} arises in in Eq.(\ref{V}) due to
the infrared (IR) divergences, which begin to manifest themselves in the static potential at the
three-loop level. However, in the concrete applications of the effective non-relativistic QCD
these IR-divergent terms are cancelling out with certain ultraviolet (UV) divergent terms originated by the interaction of ultrasoft gluons with the 
heavy quark-antiquark bound states (see e.g.  \cite{Brambilla:2004jw, Kniehl:2002br}). 
Since we consider regions of the intermediate and high energies only, these IR-corrections will not affect the behavior of various physical quantities and 
therefore we will not take them into account in our RG-oriented studies. 

The analytical expression for the one-loop coefficient 
$a_1$ in Eq.(\ref{V}) was calculated in 
Refs.\cite{Fischler:1977yf, Billoire:1979ih}, 
while two-loop one $a_2$ is known from the calculations of Refs.\cite{Peter:1996ig, Schroder:1998vy}. These terms read:
\begin{eqnarray}
\label{a1}
a_1&=& \frac{31}{36}C_A-\frac{5}{9}T_Fn_f, \\
\label{a2}
a_2&=& \bigg(\frac{4343}{2592}+\frac{\pi^2}{4}-\frac{\pi^4}{64}+\frac{11}{24}\zeta_3\bigg)C_A^2-\bigg(\frac{899}{648}+\frac{7}{6}\zeta_3\bigg)C_AT_Fn_f \\ \nonumber 
&-&\bigg(\frac{55}{48}-\zeta_3\bigg)C_FT_Fn_f+\bigg(\frac{5}{9}T_Fn_f\bigg)^2.
\end{eqnarray}

Here $n_f$ is  the flavor number of active quarks and $\zeta_n=\sum\limits_{k=1}^{\infty} k^{-n}$ is the Riemann zeta-function. 

The eigenvalues $C_F$ and $C_A$ of the quadratic Casimir operator in the fundamental
and adjoint  representation of the generic simple gauge group are defined as $(T^aT^a)_{ij}=C_F\delta_{ij}$ 
and $f^{acd}f^{bcd}=C_A\delta^{ab}$ correspondingly, 
where $T^a$ are generators of the Lie algebra of 
the considered gauge group in the fundamental representation with the 
corresponding commutation relation $[T^a, T^b]$=$if^{abc}T^c$. They are normalized as $Tr(T^aT^b)=T_F\delta^{ab}$ with the Dynkin index $T_F$. Note that in our study we are primarily interested in the case of
the $SU(N_c)$ color group with $C_A=N_c$ and $C_F=(N^2_c-1)/(2N_c)$, $T_F=1/2$,  and its particular case of the $SU(3)$-group, relevant
for physical QCD.

The three-loop contribution $a_3$ is a cubic polynomial in $n_f$:
\begin{equation}
\label{a3}
a_3=a^{(3)}_3n^3_f+a^{(2)}_3n^2_f+a^{(1)}_3n_f+a^{(0)}_3.
\end{equation}

The terms leading in powers of $n_f$ can be extracted from the renormalon-chain contributions to the Coulomb QED static 
potential (or from the representation of the QED invariant charge directly related to the photon vacuum polarization function \cite{Gorishnii:1991hw}).  
The analytical expression for the quadratic $n^2_f$-coefficient was obtained in Ref.\cite{Smirnov:2008pn}. Because of the technical difficulties, the contributions $a^{(1)}_3$ and $a^{(0)}_3$ were calculated analytically later in Ref.\cite{Lee:2016cgz}. They turned out to be much more complicated, than the coefficient $a^{(2)}_3$. Indeed,  
in addition to the expected appearance of 
$\pi^2$, $\pi^4$, $\zeta_3$, $\pi^2\zeta_3$ and $\zeta_5$-terms (see e.g. \cite{Kataev:2015yha}), 
the results of \cite{Lee:2016cgz}
also contain contributions proportional to 
$\pi^2\log 2$, $\pi^4\log 2$
and, more substantially, the basic constants with the new greatest weight of transcendence six $w=6$, namely $\pi^6$, $\zeta^2_3$,  
 $\pi^2\zeta_3\log 2$, $\pi^4\log^2 2$ 
 and the ones that include 
more complicated functions, viz $\pi^2\alpha_4$ and $s_6$, where $\alpha_4={\rm{Li}}_4(1/2)+\log^4 2/4!$ with polylogarithmic function ${\rm{Li}}_{n}(x)=\sum_{k=1}^{\infty}x^kk^{-n}$ and $s_6=\zeta_6+\zeta_{-5,-1}$ with $\zeta_6=\pi^6/945$ and multiple zeta value $\zeta_{-5,-1}=\sum_{k=1}^{\infty}\sum_{i=1}^{k-1}(-1)^{i+k}/ik^5$ (see Appendix A). 

For the convenience of readers and for the purposes of the further discussion it is useful to present all four coefficients in flavor expansion (\ref{a3}):
\begin{subequations}
\begin{eqnarray}
\label{a33}
a^{(3)}_3&=&-\bigg(\frac{5}{9}\bigg)^3T^3_F, \\
\label{a32}
a^{(2)}_3&=&\bigg(\frac{12541}{15552}+\frac{23}{12}\zeta_3+\frac{\pi^4}{135}\bigg)C_AT^2_F+\bigg(\frac{7001}{2592}-\frac{13}{6}\zeta_3\bigg)C_FT^2_F, \\
\label{a31}
a^{(1)}_3&=&\bigg[-\frac{58747}{31104}-\frac{89}{16}\zeta_3+\frac{761}{161280}\pi^6-\frac{3}{4}s_6
+\pi^4\bigg(-\frac{157}{3456}-\frac{5}{576}\log 2+\frac{\log^2 2}{64}\bigg) \\ \nonumber
&+&\pi^2\bigg(\frac{17}{1728}-\frac{19}{192}\zeta_3-\frac{\log 2}{48}-\frac{7}{32}\zeta_3\log 2-\frac{\alpha_4}{2}\bigg) 
+\frac{1091}{384}\zeta_5+\frac{57}{128}\zeta^2_3
\bigg]C^2_AT_F \\ \nonumber
&+&\bigg(-\frac{71281}{10368}+\frac{33}{8}\zeta_3+\frac{5}{4}\zeta_5\bigg)C_AC_FT_F+\bigg(\frac{143}{288}+
\frac{37}{24}\zeta_3-\frac{5}{2}\zeta_5\bigg)C^2_FT_F
+\bigg[\frac{5}{96}\pi^6 \\ \nonumber
&+&\pi^4\bigg(-\frac{23}{24}+\frac{\log 2}{6}-\frac{\log^2 2}{2}\bigg)+\pi^2\bigg(\frac{79}{36}-\frac{61}{12}\zeta_3+\log 2+\frac{21}{2}\zeta_3\log 2\bigg)\bigg]\frac{d^{abcd}_Fd^{abcd}_F}{N_A}, \\
\label{a30}
a^{(0)}_3&=&\bigg[\frac{385645}{186624}+\frac{73}{24}\zeta_3-\frac{4621}{193536}\pi^6+\frac{9}{4}s_6
+\pi^4\bigg(\frac{1349}{17280}-\frac{5}{144}\log 2-\frac{5}{72}\log^2 2\bigg) \\ \nonumber
&+&\pi^2\bigg(-\frac{953}{3456}+\frac{175}{128}\zeta_3-\frac{461}{288}\log 2+\frac{217}{192}\zeta_3\log 2+\frac{73}{24}\alpha_4\bigg)-\frac{1927}{384}\zeta_5-\frac{143}{128}\zeta^2_3\bigg]C^3_A 
\end{eqnarray}
\begin{eqnarray}
\nonumber
&+&\bigg[\frac{1511}{2880}\pi^6+\pi^4\bigg(-\frac{39}{16}+\frac{35}{12}\log 2+\frac{31}{12}\log^2 2\bigg)+\pi^2\bigg(\frac{929}{72}-\frac{827}{24}\zeta_3-74\alpha_4 \\ \nonumber
&+&\frac{461}{6}\log 2-\frac{217}{4}\zeta_3\log 2\bigg)\bigg]\frac{d^{abcd}_Fd^{abcd}_A}{N_A}.
\end{eqnarray}
\end{subequations}

Here  $d^{abcd}_F$ and $d^{abcd}_A$ are  
the rank four totally symmetric higher order group   invariants, defined in the fundamental and adjoint representations; $N_A$ is the number of  generators of the group. For the particular case of the $SU(N_c)$ gauge group the aforementioned color structures  are expressed through the number of colors $N_c$ by the following way: $N_A=N_c^2-1$,  $d_F^{abcd}d_A^{abcd}/N_A=N_c(N_c^2+6)/48$,  
$d_F^{abcd}d_F^{abcd}/N_A=(N_c^4-6N_c^2+18)/(96N_c^2)$.

Our further analysis will be in part a continuation of the work \cite{Garkusha:2018mua}, where we investigated the requirements imposed on renormalization schemes leading to the factorization of the conformal anomaly term $\beta(a_s)/a_s$ in the generalized Crewther relation
\begin{equation}
\label{GCR}
D^{NS}(a_s)C^{NS}_{Bjp}(a_s)=1+\Delta_{csb}(a_s)=1+\bigg(\frac{\beta(a_s)}{a_s}\bigg)K(a_s),
\end{equation}
which involves two RG-invariant Euclidean quantities, namely the flavor non-singlet (NS) contributions to the Adler function $D(Q^2)$ and 
to the coefficient function $C_{Bjp}(Q^2)$ of the Bjorken polarized sum rule. The first of them  is the characteristics of the $e^+e^-$ annihilation into hadrons, whereas the second one enters in theoretical expression for the Bjorken sum rule for deep inelastic scattering (DIS) of the polarized charged leptons on nucleons. Note that $a_s=a_s(\mu^2=Q^2)$ in Eq.(\ref{GCR}).

In the normalization used by us, the unity in Eq.(\ref{GCR}) corresponds to the original Crewther relation \cite{Crewther:1972kn}, derived in the Born approximation of the massless theory of strong interactions by means of application of the
operator product expansion (OPE) approach to the 
axial-vector-vector (AVV) triangle diagram in the conformal symmetry limit. 

It was discovered in Ref.\cite{Broadhurst:1993ru} that in the $\msbar$-scheme starting from $a^2_s$-terms of PT the Crewther relation is modified. In addition to unity, an extra contribution arises that turns out to be proportional to the RG $\beta$-function:
\begin{equation}
\label{Beta}
\mu^2\frac{\partial a_s(\mu^2)}{\partial \mu^2}=\beta(a_s(\mu^2))=-\sum\limits_{i\geq 0}\beta_i a^{i+2}_s(\mu^2).
\end{equation}

The renormalization procedure breaks the conformal symmetry of the massless QCD. In particular, this reflects the violation of the symmetry with respect to conformal transformations of the AVV function. The effect of this violation in Eq.(\ref{GCR}) is described by the conformal symmetry breaking term $\Delta_{csb}(a_s)$, proportional to the factor $\beta(a_s)/a_s$ and containing the polynomial $K(a_s)=\sum_{n\geq 1} K_n a^n_s$ in powers of $a_s$. This fact was discovered in the $\msbar$-scheme at the $\mathcal{O}(a^3_s)$ level in 
\cite{Broadhurst:1993ru} and confirmed at the $\mathcal{O}(a^4_s)$ level later in \cite{Baikov:2010je}. Now it is customary to call this form of the generalized Crewther relation as the Crewther--Broadhurst--Kataev (CBK) relation in the literature. It was  intensively studied from different points of view  e.g. in  works \cite{Kataev:2010du, Cvetic:2016rot, Gabadadze:2017ujx, Baikov:2022zvq}.

Recently the analog of the CBK relation was considered in the extended QCD model with arbitrary number of fermion representations at $\mathcal{O}(a^4_s)$ level in Ref.\cite{Chetyrkin:2022fqk}.
It was shown there that in this case the CBK relation remains valid as well. This fact confirms the non-accidental nature of factorization of $\beta$-function at least at $\mathcal{O}(a^4_s)$ order. Moreover, arguments presented in Refs.\cite{Gabadadze:1995ei, Crewther:1997ux, Braun:2003rp} indicate that the CBK relation will be true in the $\rm{\overline{MS}}$-scheme in QCD in all orders of PT. 
 
The natural question arises whether there do exist theoretical requirements on the choice of the ultraviolet subtraction schemes, which provide the realization of the fundamental property of the
$\beta$-factorization in the CBK relation. The results of \cite{Garkusha:2018mua, Kataev:2017oqg, Molokoedov:2020kxq} demonstrate that this feature of CBK will be implemented for a wide class of gauge-dependent momentum subtraction MOM-like schemes (such as, for instance, the mMOM-scheme \cite{vonSmekal:2009ae, Gracey:2013sca, Gracey:2014pba, Ruijl:2017eht, Zeng:2020lwi, Gracey:2022xjs})
in a linear covariant Landau gauge $\xi=0$\footnote{
The gauges $\xi=-3$ and $\xi=-1$ are the highlighted ones as well (for more detail see \cite{Garkusha:2018mua, Kataev:2017oqg, Molokoedov:2020kxq}).} at least at the $\mathcal{O}(a^4_s)$ level (and apparently in all orders of PT).
Therefore, the often prevailing opinion in the literature that the CBK-relation is valid only for the gauge-invariant MS-like schemes turned out to be incorrect. 

Since it is not obvious that the factorization of the RG $\beta$-function in the CBK relation will also be observed in some gauge-invariant schemes other than $\rm{MS}$-like ones, in this work we will study this issue on the example of the effective gauge-independent V-scheme. In this scheme the static potential of heavy quark-antiquark pair has the Coulomb-like form and all higher order corrections are absorbed in redefinition of the effective charge with a corresponding change in the scale parameter. For this aim and for the goals that will be discussed later, we will obtain the analytical expressions for $\beta$-function in the V-scheme in the four-loop approximation and also for both the Adler and the coefficient function of the Bjorken polarized sum rule in the V-scheme in the same order of PT in the case of generic simple gauge group. Thereby, we will perform the logical completion of the studies started in \cite{Kataev:2015yha} and continued later in \cite{Garkusha:2018mua}. Further we will generalize the consideration of the factorization of $\beta$-function in the CBK relation to a wide class of gauge-invariant subtraction schemes. Similar problems will also be investigated for the case of the QED. In the end, we will draw a number of conclusions on the relationship between $\beta$-function in the V-scheme in QED and the Gell-Man--Low $\Psi$-function, including a fixation of definite four-loop contributions to the static potential directly obtained in \cite{Grozin:2022umo}.

\section{$\beta$-function in the V-scheme}

Let us turn to the effective gauge-invariant V-scheme. It was first introduced in Refs.\cite{Peter:1996ig, Schroder:1998vy} and was used in modeling the smooth transition of the QCD coupling constant through the thresholds of heavy quark productions in the case when the mass corrections to the static potential are taken into account \cite{Brodsky:1999fr}. Other  
applications of the V-scheme in the perturbative QCD strudies may be found e.g. in Refs.\cite{Brodsky:1994eh, Kiselev:2002iy, Deur:2016tte, Karbstein:2018mzo, Hoque:2020qak, Afonin:2022aqt}).

Now we use the analytical results on the static potential presented in previous section to refine the semi-analytic form of
the fourth-order expression for the RG $\beta^V$-function in the generic simple gauge group, obtained previously in \cite{Kataev:2015yha} and  applied 
 to analysis of theoretical QCD ambiguities 
for $e^+e^-$ annihilation into hadrons $R$-ratio at the $\mathcal{O}(a_s^4)$-level  
in an energy region below manifestation of the left shoulder of $Z^0$-peak.

Summarizing the aforesaid, one can succinctly describe the V-scheme by the following expression of the static heavy quark-antiquark potential in the Coulomb-like form:
 \begin{equation}
\label{V-schemedef}
V(\vec{q}^{\; 2})=-4\pi C_F\frac{\alpha_{s, V}(\vec{q}^{\; 2})}{\vec{q}^{\; 2}}~,
\end{equation}
where all higher order PT corrections to $V(\vec{q}^{\; 2})$ are absorbed in the effective coupling $\alpha_{s, V}(\vec{q}^{\; 2})$ and, as was already stated, we neglect
the contribution of the three-loop IR logarithmic term in  Eq.(\ref{V}).
In accordance with technique of the effective charges (ECH),  developed in Refs.\cite{Grunberg:1982fw, Krasnikov:1981rp, Kataev:1981aw}, we define the effective V-scheme scale by means of the following relation, associated with its $\rm{\overline{MS}}$-scheme counterpart:
\begin{equation}
\mu^2_V=\mu^2\exp (a_1/\beta_0),
\end{equation} 
where $a_1$ is given by Eq.(\ref{a1}) and $\beta_0$ is the first scheme-independent coefficient \cite{Gross:1973id,  Politzer:1973fx} of the RG $\beta$-function (\ref{Beta}). Further, fixing $\vec{q}^{\; 2}=\mu^2_V$ one can finally gain the link between couplings in the V- and $\rm{\overline{MS}}$-scheme, normalized at one V-scheme scale:
\begin{equation}
\label{asV-as}
\alpha_{s, V}(\mu^2_V)=\alpha_s(\mu^2_V)\bigg(1+a_1a_s(\mu^2_V)+a_2a^2_s(\mu^2_V)+a_3a^3_s(\mu^2_V)+\mathcal{O}(a^4_s)\bigg).
\end{equation}
After that we define $\beta$-function in the V-scheme that governs the scale dependence of $\alpha_{s, V}$
\begin{equation}
\beta^V(a_{s,V})=\mu^2_V\frac{\partial a_{s,V}}{\partial\mu^2_V}=-\sum\limits_{i\geq 0} \beta^V_i a^{i+2}_{s,V}
\end{equation}
and its relation to the $\rm{\overline{MS}}$-scheme $\beta$-function:
\begin{equation}
\label{beta-relation}
\beta^V(a_{s,V}(a_s))=\beta(a_s)\frac{\partial a_{s,V}(a_s)}{\partial a_s}~
\end{equation}

The combination of the Eqs.(\ref{asV-as}) and (\ref{beta-relation}) yields the following relationships between coefficients of $\beta$-functions in the V- and $\rm{\overline{MS}}$-scheme:
\begin{subequations}
\begin{eqnarray}
\label{b1}
\beta^V_0&=&\beta_0, ~~~ 
\beta^V_1=\beta_1, \\
\label{b2}
\beta^V_2&=&\beta_2-a_1\beta_1+(a_2-a^2_1)\beta_0,
\\
\label{b3}
\beta^V_3&=&\beta_3-2a_1\beta_2+a^2_1\beta_1+(2a_3-6a_1a_2+4a^3_1)\beta_0, 
\end{eqnarray}
and even in higher orders of PT with still unknown correction $a_4$ to the static potential 
\begin{eqnarray}
\label{b4}
\beta^V_4&=&\beta_4-3a_1\beta_3+(4a^2_1-a_2)\beta_2+(a_3-2a_1a_2)\beta_1 \\ \nonumber
&+&(3a_4-12a_1a_3-5a^2_2+28a^2_1a_2-14a^4_1)\beta_0,
\end{eqnarray}
\end{subequations}
etc. These formulas reflect the transformation laws of the coefficients of the $\beta$-functions upon transition from one gauge-invariant renormalization scheme to another one. The consequence of application of the ECH approach is a scheme-invariance of all coefficients of the effective $\beta$-functions within the gauge-independent MS-like schemes (for details see \cite{Stevenson:1981vj, Kataev:1995vh}). 

The first two coefficients of $\beta^V$ coincide  identically  with their $\rm{\overline{MS}}$-analogs (\ref{b1}), calculated in \cite{Gross:1973id,  Politzer:1973fx}, \cite{Jones:1974mm, Caswell:1974gg, Egorian:1978zx} correspondingly: 
\begin{subequations}
 \begin{eqnarray}
\label{beta0}
\beta^V_0&=&\frac{11}{12}C_A-\frac{1}{3}T_Fn_f, \\ \label{beta1}
\beta^V_1&=&\frac{17}{24}C_A^2-\frac{5}{12}C_AT_Fn_f
-\frac{1}{4}C_FT_Fn_f.
\end{eqnarray}

The third and fourth terms  $\beta^V_2$ and  $\beta^V_3$ (\ref{b2}-\ref{b3})  are expressed through
three- and four-loop coefficients of the $\rm{\overline{MS}}$-scheme
RG $\beta$-function, analytically computed in  Refs.\cite{Tarasov:1980au, Larin:1993tp} and  \cite{vanRitbergen:1997va, Czakon:2004bu} respectively. Using Eqs.(\ref{a1}-\ref{a2}) and 
(\ref{b2}), one can obtain the analytic three-loop coefficient $\beta^V_2$:
\begin{eqnarray}
\label{beta2V}
\beta^V_2&=&\bigg(\frac{103}{96}+
\frac{121}{288}\zeta_3+\frac{11}{48}\pi^2-\frac{11}{768}\pi^4\bigg)C^3_A \\ \nonumber
&+&\bigg(-\frac{445}{576}-\frac{11}{9}\zeta_3-\frac{\pi^2}{12}+\frac{\pi^4}{192}\bigg)C^2_AT_Fn_f 
+\bigg(-\frac{343}{288}
+\frac{11}{12}\zeta_3\bigg)C_AC_FT_Fn_f \\ \nonumber
&+&\frac{1}{32}C^2_FT_Fn_f +
\bigg(\frac{1}{288}+\frac{7}{18}\zeta_3\bigg)
C_AT^2_Fn^2_f
+\bigg(\frac{23}{72}-\frac{1}{3}\zeta_3\bigg)C_FT^2_Fn^2_f~. 
\end{eqnarray}

This  result was originally derived  in  Ref.\cite{Schroder:1998vy}. Unlike the $\rm{\overline{MS}}$-scheme $\beta_2$-term, the coefficient $\beta^V_2$ contains not only the rational numbers but the transcendental ones as well, namely the  $\zeta_3$, $\pi^2$ and $\pi^4$-contributions. They originate from two-loop correction $a_2$ (\ref{a2}) to the static potential.

Utilizing now Eqs.(\ref{a33}-\ref{a30}) and (\ref{b3}), we find the four-loop coefficient $\beta^V_3$ in analytical form: 
\begin{align}
\label{beta3V}
&\beta^V_3=\bigg[-\frac{3871}{2592}+\frac{1463}{432}\zeta_3-\frac{21197}{2304}\zeta_5-\frac{1573}{768}\zeta^2_3+\frac{33}{8}s_6-\frac{50831}{1161216}\pi^6   \\  \nonumber
&+\pi^4\bigg(\frac{45023}{207360}-\frac{55}{864}\log 2-\frac{55}{432}\log^2 2\bigg)+\pi^2\bigg(-\frac{35035}{20736}+\frac{1925}{768}\zeta_3
+\frac{803}{144}\alpha_4-\frac{5071}{1728}\log 2 \\ \nonumber
&+\frac{2387}{1152}\zeta_3\log2\bigg)\bigg]C^4_A
+\bigg[\frac{731}{192}-\frac{13}{3}\zeta_3+\frac{19709}{2304}\zeta_5+\frac{1199}{768}\zeta^2_3-\frac{23}{8}s_6
+\frac{10189}{414720}\pi^6 \\ \nonumber
&+\pi^4\bigg(-\frac{2419}{11520}+\frac{25}{3456}\log 2 +\frac{259}{3456}\log^2 2\bigg)+\pi^2\bigg(\frac{14477}{10368}-\frac{1259}{1152}\zeta_3-\frac{53}{18}\alpha_4+\frac{889}{864}\log 2 \\ \nonumber
&-\frac{665}{576}\zeta_3\log 2\bigg)\bigg]C^3_AT_Fn_f
+\bigg[-\frac{7645}{1152}+\frac{61}{24}\zeta_3+\frac{55}{24}\zeta_5\bigg]C^2_AC_FT_Fn_f+\frac{23}{128}C^3_FT_Fn_f \\  \nonumber
&+\bigg[\frac{143}{576}+\frac{143}{48}\zeta_3-\frac{55}{12}\zeta_5\bigg]C_AC^2_FT_Fn_f+\bigg[-\frac{1171}{432}+\frac{89}{72}\zeta_3-\frac{1091}{576}\zeta_5-\frac{19}{64}\zeta^2_3+\frac{1}{2}s_6 \\ \nonumber
&-\frac{761}{241920}\pi^6+\pi^4\bigg(\frac{529}{8640}+\frac{5}{864}\log 2-\frac{1}{96}\log^2 2\bigg)+\pi^2\bigg(-\frac{737}{2592}+\frac{19}{288}\zeta_3+\frac{1}{3}\alpha_4 \\ \nonumber
&+\frac{1}{72}\log 2+\frac{7}{48}\zeta_3\log 2\bigg)\bigg]C^2_AT^2_Fn^2_f+\bigg[\frac{583}{144}-\frac{7}{4}\zeta_3-\frac{5}{6}\zeta_5\bigg]C_AC_FT^2_Fn^2_f+\bigg[-\frac{29}{288}-\frac{4}{3}\zeta_3 \\ \nonumber
&+\frac{5}{3}\zeta_5\bigg]C^2_FT^2_Fn^2_f+\bigg[\frac{293}{648}+\frac{\zeta_3}{54}-\frac{2}{405}\pi^4\bigg]C_AT^3_Fn^3_f+\bigg[-\frac{1}{2}+\frac{\zeta_3}{3}\bigg]C_FT^3_Fn^3_f+\bigg[-\frac{5}{144} \\ \nonumber
&+\frac{11}{12}\zeta_3\bigg]\frac{d^{abcd}_Ad^{abcd}_A}{N_A}+\bigg[\frac{2}{9}-\frac{13}{6}\zeta_3\bigg]\frac{d^{abcd}_Fd^{abcd}_A}{N_A}n_f+\bigg[-\frac{1511}{4320}\pi^6+\pi^4\bigg(\frac{13}{8}-\frac{35}{18}\log 2  \\ \nonumber
&-\frac{31}{18}\log^2 2\bigg)
+\pi^2\bigg(-\frac{929}{108}+\frac{827}{36}\zeta_3+\frac{148}{3}\alpha_4-\frac{461}{9}\log 2+\frac{217}{6}\zeta_3\log 2\bigg)\bigg]\frac{d^{abcd}_Fd^{abcd}_A}{N_A}T_Fn_f \\ \nonumber
&+\bigg[\frac{16621}{17280}\pi^6+\pi^4\bigg(-\frac{143}{32}+\frac{385}{72}\log 2+\frac{341}{72}\log^2 2\bigg)+\pi^2\bigg(\frac{10219}{432}-\frac{9097}{144}\zeta_3-\frac{407}{3}\alpha_4 \\ \nonumber
&+\frac{5071}{36}\log 2-\frac{2387}{24}\zeta_3\log 2\bigg)\bigg]C_A\frac{d^{abcd}_Fd^{abcd}_A}{N_A}+\bigg[\frac{55}{576}\pi^6-\pi^4\bigg(\frac{253}{144}-\frac{11}{36}\log 2+\frac{11}{12}\log^2 2\bigg) 
\end{align}
\begin{align}
 \nonumber
&+\pi^2\bigg(\frac{869}{216}-\frac{671}{72}\zeta_3+\frac{11}{6}\log 2+\frac{77}{4}\zeta_3\log 2\bigg)\bigg]C_A\frac{d^{abcd}_Fd^{abcd}_F}{N_A}n_f-\bigg[\frac{11}{36}-\frac{2}{3}\zeta_3\bigg]\frac{d^{abcd}_Fd^{abcd}_F}{N_A}n^2_f \\
\nonumber
&+\bigg[-\frac{5}{144}\pi^6+\pi^4\bigg(\frac{23}{36}-\frac{1}{9}\log 2+\frac{1}{3}\log^2 2\bigg)+\pi^2\bigg(-\frac{79}{54}+\frac{61}{18}\zeta_3-\frac{2}{3}\log 2 \\ \nonumber
&-7\zeta_3\log 2\bigg)\bigg]\frac{d^{abcd}_Fd^{abcd}_F}{N_A}T_Fn^2_f.
\end{align}
\end{subequations}

The analytical result (\ref{beta3V}) improves the presentation of our previous semi-analytic expression for $\beta^V_3$, obtained in \cite{Kataev:2015yha}. Indeed, the coefficient  $\beta^V_3$ presented there contained numerical uncertainties, associated with the inability to calculate specific three-loop master integrals to the static potential with high precision \cite{Smirnov:2008pn, Smirnov:2009fh, Anzai:2009tm} sufficient to apply the PSLQ algorithm \cite{PSLQ, Bailey:1999nv} and restore their analytical expressions from the obtained numerical values. This problem was solved in Ref.\cite{Lee:2016cgz} by means of the dimensional recurrence relation \cite{Lee:2009dh} and the convergence acceleration algorithm \cite{Lee:2015eva}. 

The expression (\ref{beta3V}) is rather cumbersome: unlike $\beta_3$-coefficient in the $\rm{\overline{MS}}$-scheme, which contains rational numbers and $\zeta_3$-contributions only, the coefficient $\beta^V_3$ is expressed through a much larger number of terms with higher transcendentalities initially appearing in the three-loop correction $a_3$ to the static potential. Note also that the result (\ref{beta3V}) includes four extra color structures originating from $2a_3\beta_0$-term in Eq.(\ref{b3}) and 
 not encountered in the representation of $\beta_3$-coefficient, viz $C_Ad^{abcd}_Fd^{abcd}_A/N_A$, $C_Ad^{abcd}_Fd^{abcd}_Fn_f/N_A$, $d^{abcd}_Fd^{abcd}_AT_Fn_f/N_A$ and $d^{abcd}_Fd^{abcd}_FT_Fn^2_f/N_A$ patterns. The term proportional to the  $d^{abcd}_Ad^{abcd}_A/N_A$-structure in (\ref{beta3V}) follows from the $\rm{\overline{MS}}$-scheme coefficient $\beta_3$. For the particular case of the $SU(N_c)$ gauge group, the discussed term is equal to  $d^{abcd}_Ad^{abcd}_A/N_A=N^2_c(N^2_c+36)/24$.

Taking into account the values $\alpha_4\approx 0.5270972$, $s_6\approx 0.9874414$, we arrive to the following numerical form of Eqs.(\ref{beta0}-\ref{beta3V}) in the case of the $SU(3)$ color gauge group:
\begin{subequations}
\begin{eqnarray}
\beta^V_0 &=& 2.75-0.1666667n_f,  \\ 
\beta^V_1 &=& 6.375-0.7916667n_f,  \\
\label{b2Vnum}
\beta^V_2 &=& 66.00284-11.656347n_f+0.3261237n_f^2, \\ 
\label{b3Vnum}
\beta^V_3 &=& 168.6484-50.59222n_f+2.761578n^2_f-0.0190318n^3_f.
\end{eqnarray}
\end{subequations}

The expressions (\ref{b2Vnum}-\ref{b3Vnum}) should be compared with their $\msbar$ counterparts \cite{Tarasov:1980au, Larin:1993tp},  \cite{vanRitbergen:1997va, Czakon:2004bu}
\begin{subequations}
\begin{eqnarray}
\label{b2MSnum}
\beta_2 &=& 22.32031-4.368924n_f+0.0940394n^2_f, \\
\label{b3MSnum}
\beta_3 &=& 114.2303-27.13394n_f+1.582379n^2_f+0.0058567n^2_f,
\end{eqnarray}
\end{subequations}
and with the mMOM ones in the Landau gauge \cite{vonSmekal:2009ae, Gracey:2013sca, Ruijl:2017eht}
\begin{subequations}
\begin{eqnarray}
\label{b2mMOMnum}
\beta^{{\rm{mMOM}}, \;\xi=0}_2 &=&
47.50754 - 9.771667n_f + 0.3028642n^2_f, \\
\label{b3mMOMnum}
\beta^{{\rm{mMOM}}, \;\xi=0}_3 &=& 
392.7385 - 95.40363n_f + 6.349228n^2_f - 0.1073931n^3_f.
\end{eqnarray}
\end{subequations}

Naturally, the first two coefficients of the RG $\beta$-function in the V-, $\msbar$- and mMOM-scheme in the Landau gauge coincide respectively.

\section{The Adler function, $R$-ratio and the Bjorken polarized sum rule in the V-scheme}

\subsection{The Adler function in the V-scheme}

As known, the Adler function $D(Q^2)$ is the convenient ingredient for calculating the Minkowskian annihilation electron-positron cross section into hadrons with help of the K\"allen--Lehmann-type dispersion relation (see e.g. \cite{Nesterenko:2017wpb, Davier:2023hhn}). It is determined in the Euclidean domain with the Euclidean transferred momentum $Q^2=-q^2$ and, what is very substantial, is a renorm-invariant quantity. Its two-, three- and four-loop expressions in the $\rm{\overline{MS}}$-scheme were directly  evaluated in Refs.\cite{Chetyrkin:1979bj, Dine:1979qh, Celmaster:1979xr}, \cite{Gorishnii:1990vf, Surguladze:1990tg} and \cite{Baikov:2008jh, Baikov:2010je, Baikov:2012zn} respectively. 

In the massless limit the Adler function is decomposed into a sum of the flavor non-singlet (NS) and singlet (SI) components:
\begin{equation}
\label{NS-SI-Adler}
D(a_s)=d_R\bigg(\sum\limits_f Q^2_f\bigg) D^{NS}(a_s)+d_R\bigg(\sum\limits_f Q_f\bigg)^2 D^{SI}(a_s),
\end{equation}
where $d_R$ is the dimension of the quark representation of the Lie algebra of the considered generic simple gauge group. In the case of the $SU(N_c)$ color gauge group $d_R=N_c$. $Q_f$ is the electric charge of the active quark
with flavor $f$. The singlet (SI) flavor contribution $D^{SI}(a_s)$ appears from the third order of PT due to the special diagrams of the light-by-light scattering type \cite{Gorishnii:1990vf, Baikov:2012zn}.

In order to obtain the analytic four-loop expression for the NS Adler function $D^{NS}_V(a_{s, V})$ in the V-scheme, we use its explicit $\msbar$-scheme result at the $\mathcal{O}(a^4_s)$ level, the relation (\ref{asV-as}) between the corresponding couplings in
two considered gauge-independent schemes and take into account that the flavor NS Adler function is the RG-invariant quantity. Keeping in mind the aforesaid, we get the following results:
\begin{subequations}
\begin{eqnarray}
&&D^{NS}_V(a_{s,V})=1+\sum\limits_{k\geq 1} d^{NS}_{k, V} a^k_{s, V}, \\ 
\label{d1V}
d^{NS}_{1, V}&=&\frac{3}{4}C_F, \\ 
\label{d2V}
d^{NS}_{2, V}&=&-\frac{3}{32}C^2_F+\bigg(\frac{307}{96}-\frac{11}{4}\zeta_3\bigg)C_FC_A+\bigg(-\frac{23}{24}+\zeta_3\bigg)C_FT_Fn_f, \\
\label{d3V}
d^{NS}_{3, V}&=&-\frac{69}{128}C^3_F+\bigg(-\frac{175}{96}-\frac{143}{16}\zeta_3+\frac{55}{4}\zeta_5\bigg)C^2_FC_A 
\\ \nonumber
&+&\bigg(\frac{621}{32}-\frac{1403}{96}\zeta_3-\frac{55}{24}\zeta_5-\frac{3}{16}\pi^2+\frac{3}{256}\pi^4\bigg)C_FC^2_A+\bigg(\frac{3}{2}-\zeta_3\bigg)C_FT^2_Fn^2_f \\ \nonumber
&+&\bigg(-\frac{375}{32}+\frac{205}{24}\zeta_3+\frac{5}{6}\zeta_5\bigg)C_FC_AT_Fn_f+\bigg(\frac{29}{96}+4\zeta_3-5\zeta_5\bigg)C^2_FT_Fn_f, \\
\label{d4V}
d^{NS}_{4, V}&=&\bigg(\frac{4157}{2048}+\frac{3}{8}\zeta_3\bigg)C^4_F+\bigg(-\frac{3335}{512}-\frac{139}{128}\zeta_3+\frac{2255}{32}\zeta_5-\frac{1155}{16}\zeta_7\bigg)C^3_FC_A \\ \nonumber
&+&\bigg(-\frac{498269}{18432}-\frac{17513}{192}\zeta_3+100\zeta_5+\frac{1155}{32}\zeta_7+\frac{3}{64}\pi^2-\frac{3}{1024}\pi^4\bigg)C^2_FC^2_A 
\end{eqnarray}
\begin{eqnarray}
\nonumber 
&+&\bigg[\frac{668335}{4608}-\frac{101621}{1152}\zeta_3-\frac{89119}{1536}\zeta_5+\frac{34199}{1536}\zeta^2_3-\frac{385}{64}\zeta_7-\frac{27}{16}s_6+\frac{4621}{258048}\pi^6 \\ \nonumber 
&+&\pi^4\bigg(\frac{1}{90}-\frac{11}{128}\zeta_3+\frac{5}{192}\log 2+\frac{5}{96}\log^2 2\bigg)+\pi^2\bigg(-\frac{4183}{4608}+\frac{179}{512}\zeta_3-\frac{73}{32}\alpha_4 \\ \nonumber
&+&\frac{461}{384}\log 2-\frac{217}{256}\zeta_3\log 2\bigg)\bigg]C_FC^3_A+\bigg(\frac{287}{256}+\frac{17}{8}\zeta_3-\frac{235}{8}\zeta_5+\frac{105}{4}\zeta_7\bigg)C^3_FT_Fn_f \\ \nonumber
&+&\bigg(\frac{12277}{1152}+\frac{1117}{16}\zeta_3-\frac{145}{2}\zeta_5-\frac{11}{4}\zeta^2_3-\frac{105}{8}\zeta_7\bigg)C^2_FC_AT_Fn_f \\
 \nonumber
&+&\bigg[-\frac{201725}{1536}+\frac{41071}{576}\zeta_3+\frac{87847}{1536}\zeta_5-\frac{20225}{1536}\zeta^2_3+\frac{35}{16}\zeta_7+\frac{9}{16}s_6-\frac{761}{215040}\pi^6 \\ \nonumber
&+&\pi^4\bigg(\frac{109}{4608}+\frac{\zeta_3}{32}+\frac{5}{768}\log 2-\frac{3}{256}\log^2 2\bigg)+\pi^2\bigg(\frac{367}{2304}-\frac{109}{256}\zeta_3+\frac{3}{8}\alpha_4+\frac{\log 2}{64} \\ \nonumber
&+&\frac{21}{128}\zeta_3\log 2\bigg)\bigg]C_FC^2_AT_Fn_f+\bigg(-\frac{125}{384}-\frac{281}{24}\zeta_3+\frac{25}{2}\zeta_5+\zeta^2_3\bigg)C^2_FT^2_Fn^2_f \\ \nonumber
&+&\bigg(\frac{81103}{2304}-\frac{4859}{288}\zeta_3-\frac{35}{2}\zeta_5+\frac{11}{6}\zeta^2_3-\frac{\pi^4}{180}\bigg)C_FC_AT^2_Fn^2_f-\bigg(\frac{67}{24}-\frac{7}{6}\zeta_3-\frac{5}{3}\zeta_5\bigg)C_FT^3_Fn^3_f \\ \nonumber
&+&\bigg(\frac{3}{16}-\frac{\zeta_3}{4}-\frac{5}{4}\zeta_5\bigg)\frac{d^{abcd}_Fd^{abcd}_A}{d_R}+\bigg(-\frac{13}{16}-\zeta_3+\frac{5}{2}\zeta_5\bigg)\frac{d^{abcd}_Fd^{abcd}_F}{d_R}n_f \\ \nonumber
&+&\bigg[-\frac{1511}{3840}\pi^6+\pi^4\bigg(\frac{117}{64}-\frac{35}{16}\log 2-\frac{31}{16}\log^2 2\bigg)+\pi^2\bigg(-\frac{929}{96}+\frac{827}{32}\zeta_3+\frac{111}{2}\alpha_4 \\ \nonumber
&-&\frac{461}{8}\log 2+\frac{651}{16}\zeta_3\log 2\bigg)\bigg]C_F\frac{d^{abcd}_Fd^{abcd}_A}{N_A}+\bigg[-\frac{5}{128}\pi^6+\pi^4\bigg(\frac{23}{32}-\frac{\log 2}{8}+\frac{3}{8}\log^2 2\bigg) \\ \nonumber
&+&\pi^2\bigg(-\frac{79}{48}+\frac{61}{16}\zeta_3-\frac{3}{4}\log 2-\frac{63}{8}\zeta_3\log 2\bigg)\bigg]C_F\frac{d^{abcd}_Fd^{abcd}_F}{N_A}n_f.
\end{eqnarray}
\end{subequations}

Let us make a few comments on the derived expressions. First of all, unlike $\rm{\overline{MS}}$-scheme three-loop results, the coefficient $d^{NS}_{3, V}$ (\ref{d3V}) contains the
complementary terms into $C_FC^2_A$-contribution, which are proportional to $\pi^2$ and $\pi^4$. Secondly, the coefficient $d^{NS}_{4, V}$  (\ref{d4V})  
includes all transcendental basic constants, comprised
in $\beta^V_3$, plus the extra $\zeta_7$-term with the greatest transcendence of weight $w=7$, originally appearing from $d_4$ in the $\rm{\overline{MS}}$-scheme \cite{Baikov:2008jh}. It is also worth noting that in contrast to $d_4$, the analytic expression for $d^{NS}_{4, V}$ has two additional color structures, namely $C_Fd^{abcd}_Fd^{abcd}_A/N_A$ and $C_Fd^{abcd}_Fd^{abcd}_Fn_f/N_A$, coming from the product of $a_3$ on $d_1$.

For the $SU(3)$ case the numerical form of these coefficients read:
\begin{subequations}
\begin{eqnarray}
\label{dNS1-num}
d^{NS}_{1, V}&=&1, \\
\label{dNS2-num}
d^{NS}_{2, V}&=&-0.597626+0.1624824n_f, \\ 
\label{dNS3-num}
d^{NS}_{3, V}&=&-7.21638-1.240217n_f+0.0993144n_f^2,\\
\label{dNS4-num}
d^{NS}_{4, V}&=&19.9437+4.38696n_f- 1.114839n_f^2+0.0564909n_f^3.
\end{eqnarray}
\end{subequations}

The SI contributions to the coefficients $d_3$ and $d_4$ of the Adler function 
were calculated in the $\rm{\overline{MS}}$-scheme in Refs.\cite{Gorishnii:1990vf, Baikov:2012zn} correspondingly. Taking into account the renormalization invariance of the function $D(Q^2)$, one can obtain the SI contributions to the coefficients $d_{3, V}$ and $d_{4, V}$:
\begin{subequations}
\begin{eqnarray}
&&D^{SI}_V(a_{s, V})=\sum\limits_{k\geq 3} d^{SI}_{k, V}a^k_{s, V}, \\ 
\label{d3SI}
d^{SI}_{3, V}&=&d^{SI}_3 =\bigg(\frac{11}{192}-\frac{\zeta_3}{8}\bigg)\frac{d^{abc}d^{abc}}{d_R}~, \\ 
\label{d4SI}
d^{SI}_{4, V}&=&d^{SI}_4-3a_1d^{SI}_3=\bigg[\bigg(-\frac{13}{64}-\frac{\zeta_3}{4}+\frac{5}{8}\zeta_5\bigg)C_F+\bigg(\frac{3211}{4608}-\frac{383}{384}\zeta_3 \\ \nonumber
&+&\frac{45}{64}\zeta_5-\frac{11}{32}\zeta^2_3\bigg)C_A +\bigg(-\frac{47}{288}+\frac{19}{96}\zeta_3-\frac{5}{16}\zeta_5+\frac{\zeta^2_3}{8}\bigg)T_Fn_f\bigg]\frac{d^{abc}d^{abc}}{d_R},
\end{eqnarray}
\end{subequations}
where $d^{abc}$ is the symmetric color constant, which for the case of the $SU(N_c)$-group, we are interested in, obeys the relation $d^{abc}d^{abc}=(N^2_c-4)(N^2_c-1)/N_c$. 

The expressions (\ref{d3SI}) and (\ref{d4SI}) in the numerical form are:
\begin{equation}
\label{d3SIVnum}
d^{SI}_{3, V}=-0.413179, ~~~~ d^{SI}_{4, V}=-2.74010-0.152688n_f.
\end{equation}

\subsection{The $R(s)$-ratio in the V-scheme}

Let us move on to the case of $R(s)$-ratio of process of the electron-positron annihilation into hadrons. This quantity is directly measured in the Minkowski region of energies and is expressed through the cross section of this process:
\begin{eqnarray}
R(s)=\frac{\sigma(e^+e^-\rightarrow \gamma^*\rightarrow \text{hadrons})}{\sigma_{Born}(e^+e^-\rightarrow\gamma^*\rightarrow \mu^+\mu^-)}=d_R\bigg(\sum\limits_f Q^2_f\bigg) R^{NS}(a_s)+d_R\bigg(\sum\limits_f Q_f\bigg)^2 R^{SI}(a_s),
\end{eqnarray}
where $\sigma_{Born}(e^+e^-\rightarrow\mu^+\mu^-)=4\pi\alpha^2_{EM}/3s$ is the Born massless normalization factor.

The K\"allen--Lehmann-type dispersion representation (see e.g. \cite{Nesterenko:2017wpb, Davier:2023hhn}), related the Adler function to $R(s)$-ratio, dictates the following  analytic correspondence:
\begin{equation}
\label{R-D}
R(s)=D(s)-\frac{\pi^2}{3}d_1\beta^2_0a^3_s-\pi^2\bigg(d_2\beta^2_0+\frac{5}{6}d_1\beta_1\beta_0\bigg)a^4_s+\mathcal{O}(a^5_s)~.
\end{equation}

The terms proportional to $\pi^2$ appear here as an effect of the  analytic continuation from the Euclidean to Minkowskian domain. 

Using Eq.(\ref{R-D}) and reckoning for the RG invariance of the $R$-ratio, one can conclude that there are valid the following relations between the coefficients of the NS and SI contributions to the $R$-ratio and the Adler function in the V-scheme:
\begin{subequations}
\begin{eqnarray}
R^{NS}_V(a_{s,V})&=&1+\sum\limits_{k\geq 1} r^{NS}_{k, V} a^k_{s, V}, ~~~~ R^{SI}_V(a_{s,V})=\sum\limits_{k\geq 3} r^{SI}_{k, V} a^k_{s, V},\\
\label{rd1V}
r^{NS}_{1, V}&=&d^{NS}_{1, V}, ~~~~~ r^{NS}_{2, V}=d^{NS}_{2, V}, \\
\label{rd3V}
r^{NS}_{3, V}&=&d^{NS}_{3, V}-\frac{\pi^2}{3}d_1\beta^2_0, ~~~~~ r^{SI}_{3, V}=d^{SI}_{3, V}, \\
\label{rd4V}
r^{NS}_{4, V}&=&d^{NS}_{4, V}-\pi^2\bigg(d^{NS}_{2, V}\beta^2_0+\frac{5}{6}d_1\beta_1\beta_0\bigg), ~~~~~ r^{SI}_{4, V}=d^{SI}_{4, V}.
\end{eqnarray}
\end{subequations}

Taking into account Eqs.(\ref{rd1V}-\ref{rd4V}), we arrive at the complete numerical result for $R_V(s)$ in the V-scheme for physically relevant case of $SU(3)$ group:
\begin{subequations}
\begin{eqnarray}
R_V(a_{s, V})&=&3\sum\limits_f Q^2_f\bigg(1+\sum\limits_{k\geq 1} r_{k, V}a^{k}_{s, V}\bigg), \\
\label{r1V}
r_{1, V}&=&1, \\
r_{2, V}&=&-0.597626+0.1624824n_f, \\
r_{3, V}&=&-32.09600+1.775495n_f+0.0079291n^2_f-0.413179\delta_f, \\
\label{r4V}
r_{4, V}&=&-79.6389+13.49715n_f-0.566196n^2_f+0.0119455n^3_f  \\ \nonumber
&+&(-2.74010-0.152688n_f)\delta_f,
\end{eqnarray}
\end{subequations}
where terms with $\delta_f=(\sum_f Q_f)^2/(\sum_f Q^2_f)$ are the SI-contributions.
Note that the analogous V-scheme numerical expressions for $R$-ratio coefficients were presented previously in Ref.\cite{Kataev:2015yha} but with corresponding theoretical mean square uncertainties 
ensuing from the inaccuracies of the calculation of $a_3$-term to the static potential \cite{Smirnov:2008pn, Smirnov:2009fh, Anzai:2009tm}. Naturally, these results are in full agreement with those given in (\ref{r1V}-\ref{r4V}). The discussed uncertainties are negligible and much smaller than the ones related with the determination of physical parameters such as $\alpha_s(M^2_Z)$ \cite{Workman:2022ynf}. The interested reader may find outcomes of the study of the scheme- and scale-dependence of the $R(s)$-ratio at the $\rm{NLO}$, $\rm{NNLO}$ and $\rm{N^3LO}$ approximations in cases when $n_f=4$ and $n_f=5$ in Refs.\cite{Kataev:2015yha, Gracey:2014pba, Molokoedov:2020kxq}. 

\subsection{The Bjorken polarized sum rule in the V-scheme}

One of the important physical quantities upon investigation of the DIS process of the polarized leptons on nucleons is the coefficient Bjorken function $C_{Bjp}(Q^2)$, defining the magnitude of scaling violation in QCD. It is determined in the Euclidean region of energies  and included in the Bjorken polarized sum rule (neglecting the $\mathcal{O}(1/Q^{2k})$ nonperturbative terms):
\begin{eqnarray}
\int\limits_0^1\bigg(g^{lp}_1(x, Q^2)-g^{ln}_1(x, Q^2)\bigg)dx=\frac{1}{6}\bigg\vert\frac{g_A}{g_V}\bigg\vert C_{Bjp}(Q^2)~.
\end{eqnarray}

Here $g^{lp}_1(x, Q^2)$ and $g^{ln}_1(x, Q^2)$ are the 
structure functions of the DIS processes, which characterize the spin distribution of quarks and gluons inside nucleons, $g_A$ and $g_V$ are the axial and 
vector neutron $\beta$-decay constants with $g_A/g_V=-1.2754\pm 0.0013$ \cite{Workman:2022ynf}. 

The coefficient Bjorken function is separated in two components, namely in the NS and SI ones:
\begin{eqnarray}
C_{Bjp}(a_s)=C^{NS}_{Bjp}(a_s)+d_R\sum\limits_f Q_f C^{SI}_{Bjp}(a_s).
\end{eqnarray}

The one-, two-, three- and four-loop results for the  NS coefficient Bjorken function in the $\rm{\overline{MS}}$-scheme were obtained in Refs.\cite{Kodaira:1978sh}, \cite{Gorishnii:1985xm}, \cite{Larin:1991tj}, \cite{Baikov:2010je} correspondingly. Unlike $D^{SI}(a_s)$-function, the SI part to $C_{Bjp}(a_s)$ appears first at the $\mathcal{O}(a^4_s)$ level \cite{Larin:2013yba} and was calculated analytically in Ref.\cite{Baikov:2015tea}.

Using the explicit fourth-order approximation for the Bjorken function in the $\rm{\overline{MS}}$-scheme \cite{Baikov:2010je, Baikov:2015tea}, the  relation (\ref{asV-as}) and bearing in mind the RG-invariance of the $C_{Bjp}(a_s)$, we gain the following expressions for the NS and SI-contributions to the coefficient Bjorken function in the V-scheme:
\begin{subequations}
\begin{eqnarray}
&&C^{NS}_{Bjp, V}(a_{s, V})=1+\sum\limits_{k\geq 1} c^{NS}_{k, V}a^k_{s, V}, ~~~~~ C^{SI}_V(a_{s,V})=\sum\limits_{k\geq 4} c^{SI}_{k, V} a^k_{s, V},\\ 
\label{c1V}
c^{NS}_{1, V}&=&-\frac{3}{4}C_F, \\
\label{c2V}
c^{NS}_{2, V}&=&\frac{21}{32}C^2_F-\frac{19}{24}C_FC_A+\frac{1}{12}C_FT_Fn_f, \\
\label{c3V}
c^{NS}_{3, V}&=&-\frac{3}{128}C^3_F+\bigg(\frac{295}{288}-\frac{11}{12}\zeta_3\bigg)C^2_FC_A+\bigg(\frac{73}{36}-\frac{\zeta_3}{8}-\frac{5}{6}\zeta_5\bigg)C_FC_AT_Fn_f \\ \nonumber
&+&\bigg(-\frac{4231}{1152}+\frac{11}{32}\zeta_3+\frac{55}{24}\zeta_5+\frac{3}{16}\pi^2-\frac{3}{256}\pi^4\bigg)C_FC^2_A-\frac{5}{24}C_FT^2_Fn^2_f \\ \nonumber
&+&\bigg(-\frac{13}{36}+\frac{\zeta_3}{3}\bigg)C^2_FT_Fn_f, \\
\label{c4V}
c^{NS}_{4, V}&=&\bigg(-\frac{4823}{2048}-\frac{3}{8}\zeta_3\bigg)C^4_F+\bigg(-\frac{857}{1152}-\frac{971}{96}\zeta_3+\frac{1045}{48}\zeta_5\bigg)C^3_FC_A \\ \nonumber
&+&\bigg(\frac{776809}{55296}+\frac{4921}{384}\zeta_3-\frac{1375}{144}\zeta_5-\frac{385}{16}\zeta_7-\frac{21}{64}\pi^2+\frac{21}{1024}\pi^4\bigg)C^2_FC^2_A \\ \nonumber
&+&\bigg[-\frac{247307}{13824}+\frac{4579}{1152}\zeta_3+\frac{5557}{4608}\zeta_5-\frac{3223}{1536}\zeta^2_3+\frac{385}{64}\zeta_7+\frac{27}{16}s_6-\frac{4621}{258048}\pi^6 \\ \nonumber
&+&\pi^4\bigg(\frac{2953}{46080}-\frac{5}{192}\log 2-\frac{5}{96}\log^2 2\bigg)+\pi^2\bigg(-\frac{1361}{4608}+\frac{525}{512}\zeta_3+\frac{73}{32}\alpha_4-\frac{461}{384}\log 2 \\ \nonumber
&+&\frac{217}{256}\zeta_3\log 2\bigg)\bigg]C_FC^3_A+\bigg(\frac{317}{144}+\frac{109}{24}\zeta_3-\frac{95}{12}\zeta_5\bigg)C^3_FT_Fn_f+\bigg(-\frac{14177}{1728}-\frac{739}{144}\zeta_3 \\ \nonumber
&+&\frac{205}{72}\zeta_5+\frac{35}{4}\zeta_7\bigg)C^2_FC_AT_Fn_f+\bigg[\frac{47693}{3456}-\frac{77}{18}\zeta_3+\frac{851}{512}\zeta_5+\frac{1921}{1536}\zeta^2_3-\frac{35}{16}\zeta_7-\frac{9}{16}s_6 \\ \nonumber
&+&\frac{761}{215040}\pi^6+\pi^4\bigg(-\frac{235}{4608}-\frac{5}{768}\log 2+\frac{3}{256}\log^2 2\bigg)+\pi^2\bigg(\frac{641}{2304}-\frac{19}{256}\zeta_3-\frac{3}{8}\alpha_4 \\ \nonumber
&-&\frac{\log 2}{64}-\frac{21}{128}\zeta_3\log 2\bigg)\bigg]C_FC^2_AT_Fn_f+\bigg(\frac{1891}{3456}-\frac{\zeta_3}{36}\bigg)C^2_FT^2_Fn^2_f+\bigg(-\frac{8309}{3456}+\frac{9}{8}\zeta_3 \\ \nonumber
&-&\frac{35}{36}\zeta_5-\frac{\zeta^2_3}{6}+\frac{\pi^4}{180}\bigg)C_FC_AT^2_Fn^2_f+\frac{5}{72}C_FT^3_Fn^3_f 
+\bigg(-\frac{3}{16}+\frac{\zeta_3}{4}+\frac{5}{4}\zeta_5\bigg)\frac{d^{abcd}_Fd^{abcd}_A}{d_R} \\ \nonumber
&+&\bigg(\frac{13}{16}+\zeta_3-\frac{5}{2}\zeta_5\bigg)\frac{d^{abcd}_Fd^{abcd}_F}{d_R}n_f 
+\bigg[\frac{1511}{3840}\pi^6+\pi^4\bigg(-\frac{117}{64}+\frac{35}{16}\log 2+\frac{31}{16}\log^2 2\bigg) 
\\ \nonumber
&+&\pi^2\bigg(\frac{929}{96}-\frac{827}{32}\zeta_3-\frac{111}{2}\alpha_4+\frac{461}{8}\log 2-\frac{651}{16}\zeta_3\log 2\bigg)\bigg]C_F\frac{d^{abcd}_Fd^{abcd}_A}{N_A}+\bigg[\frac{5}{128}\pi^6 
\\ \nonumber
&-&\pi^4\bigg(\frac{23}{32}-\frac{\log 2}{8}+\frac{3}{8}\log^2 2\bigg)+\pi^2\bigg(\frac{79}{48}-\frac{61}{16}\zeta_3+\frac{3}{4}\log 2+\frac{63}{8}\zeta_3\log 2\bigg)\bigg]C_F\frac{d^{abcd}_Fd^{abcd}_F}{N_A}n_f, \\
\label{cSIV}
c^{SI}_{4, V}&=&c^{SI}_4=\frac{1}{9}\beta_0d^{abc}d^{abc}.
\end{eqnarray}
\end{subequations}

Comparing now Eqs.(\ref{d4V}) and (\ref{c4V}),  we can reveal certain similarities between analytic expressions for $d^{NS}_{4, V}$ and $c^{NS}_{4, V}$.
For instance, the contributions proportional to  $d^{abcd}_Fd^{abcd}_A/d_R$, $d^{abcd}_Fd^{abcd}_Fn_f/d_R$, $C_Fd^{abcd}_Fd^{abcd}_A/N_A$, $C_Fd^{abcd}_Fd^{abcd}_Fn_f/N_A$ color structures in their expressions are identical in absolute values, but opposite in sign. Moreover, the terms proportional to $\pi^6$, $s_6$, $\pi^4\log 2$, $\pi^4\log^2 2$, $\pi^2\alpha_4$, $\pi^2\log 2$ and $\pi^2\zeta_3\log 2$ possess the same property. Therefore,  all these color structures and transcendental constants are canceled out automatically in the sum of $d^{NS}_{4, V}+c^{NS}_{4, V}$. This fact will turn out to be important upon studying conditions of the $\beta$-function factorization in the CBK relation in the V-scheme (see discussions below). 

In the case of the $SU(3)$ group we obtain the following numerical form of the PT coefficients of $C_{Bjp, V}$-function:
\begin{subequations}
\begin{eqnarray}
&&C_{Bjp, V}(a_{s, V})=1+\sum\limits_{k\geq 1}
 c_{k, V}a^k_{s, V}, \\
 \label{c1-num}
c_{1, V}&=&-1, \\
\label{c2-num}
c_{2, V}&=&-2+0.0555556n_f, \\
\label{c3-num}
c_{3, V}&=&-2.55978+2.062006n_f-0.0694444n_f^2, \\
\label{c4-num}
c_{4, V}&=&-122.1910+30.87144n_f-1.531353n_f^2+0.0115741n_f^3 \\ \nonumber
&+&(12.22222-0.740741n_f)\eta_f,
\end{eqnarray}
\end{subequations}
where $\eta_f=\sum\limits_f Q_f$.

\subsection{PT series for the Adler function, $R$-ratio and the coefficient Bjorken function in the $\rm{\overline{MS}}$-, V- and Landau mMOM-schemes}

For comparison of the behavior of the PT series for the Adler and the coefficient Bjorken functions, we consider their expressions in the $\rm{\overline{MS}}$- \cite{Chetyrkin:1979bj, Dine:1979qh, Celmaster:1979xr, Gorishnii:1990vf, Surguladze:1990tg, Baikov:2008jh, Baikov:2010je, Baikov:2012zn}, \cite{Kodaira:1978sh, Gorishnii:1985xm, Larin:1991tj, Baikov:2015tea}, V- (\ref{dNS2-num}-\ref{dNS4-num}), (\ref{d3SIVnum}), (\ref{c2-num}-\ref{c4-num}), \cite{Molokoedov:2020kxq}  and mMOM-scheme in the Landau gauge \cite{vonSmekal:2009ae, Gracey:2014pba, Garkusha:2018mua} in the case of the $SU(3)$ color gauge group. Taking the results of the quoted works into account, we can present them in the form of Table \ref{T-1}.

\begin{table}[h]
\renewcommand{\arraystretch}{1.2}
\centering
\begin{tabular}{|m{0.08\textwidth}<{\centering}|m{0.85\textwidth}<{\centering}|}
\hline
\text{Scheme}                                                              & \textbf{The Adler function}                            \\ \hline
$\msbar$                                                                  & \begin{tabular}[c]{@{\!\!\!\!}c@{}} $1+a_s+(1.9857-0.11529n_f)a^2_s+(18.2427-4.2158n_f+0.0862n^2_f-0.413\delta_f)a^3_s$ \\$+\;(135.792-34.440n_f+1.875n^2_f-0.010n^3_f+\delta_f(-5.942+0.1916n_f))a^4_s$
\end{tabular} \\ \hline
V                                                                   & \begin{tabular}[c]{@{}c@{}}  $\!1+a_{s, V}+(-0.5976+0.1625n_f)a^2_{s, V} 
+(-7.2164-1.240n_f+0.0993n^2_f-0.413\delta_f)a^3_{s, V}$ \\
$+\;(19.944+4.387n_f-1.115n^2_f+0.056n^3_f+\delta_f(-2.740-0.1527n_f))a^4_{s, V}$
\end{tabular} \\ \hline
\begin{tabular}[c]{@{}c@{}}mMOM\\ $\xi=0$\end{tabular} & \begin{tabular}[c]{@{}c@{}}$\!\!1+a_{s, M}+(-1.535+0.1625n_f)a^2_{s, M}+(-0.6647-1.685n_f+0.0993n^2_f-0.413\delta_f)a^3_{s, M}$ \\
$+\;(-38.363+18.44n_f-1.71n^2_f+0.056n^3_f+\delta_f(-1.578-0.1527n_f))a^4_{s, M}$
\end{tabular} \\ \hline
                                                                    & \textbf{The $R$-ratio}                                 \\ \hline
$\msbar$                                                                  & \begin{tabular}[c]{@{}c@{}} $1+a_s+(1.9857-0.11529n_f)a^2_s+(-6.6369-1.2000n_f-0.00518n^2_f-0.413\delta_f)a^3_s$ \\
$+\; (-156.608+18.7748n_f-0.7974n^2_f+0.0215n^3_f   +\delta_f(-5.942+0.1916n_f))a^4_s$\end{tabular} \\ \hline
V                                                                   & \begin{tabular}[c]{@{}c@{}} $\!1+a_{s, V}+(-0.5976+0.1625n_f)a^2_{s, V}+(-32.096+1.775n_f+0.0079n^2_f-0.413\delta_f)a^3_{s, V}$\\
 $+\;(-79.639+13.497n_f-0.566n^2_f+0.0119n^3_f +
\delta_f(-2.740-0.1527n_f))a^4_{s, V}$
 \end{tabular} \\ \hline
\begin{tabular}[c]{@{}c@{}}mMOM\\ $\xi=0$\end{tabular} & \begin{tabular}[c]{@{}c@{}} $\!1+a_{s, M}+(-1.535+0.1625n_f)a^2_{s, M}+(-25.544 + 1.331n_f+0.0079n^2_f-0.413\delta_f)a^3_{s, M}$ \\
$+\;(-67.981+19.068n_f-0.904n^2_f+0.0114n^3_f
+\delta_f(-1.578-0.1527n_f))a^4_{s, M}$ 
\end{tabular} \\ \hline
                                                                    & \textbf{The coefficient Bjorken function}              \\ \hline
$\msbar$                                                                  & \begin{tabular}[c]{@{}c@{}}  $1-a_s+(-4.5833+0.33333n_f)a^2_s+(-41.4399+7.6073n_f-0.1775n^2_f)a^3_s$ \\
$+\;(-479.448+123.39n_f-7.69n^2_f+0.104n^3_f+\eta_f(12.222-0.7407n_f))a^4_s$
\end{tabular} \\ \hline
V                                                                   & \begin{tabular}[c]{@{}c@{}}  $1-a_{s, V}+(-2+0.05556n_f)a^2_{s, V}+(-2.5598+2.062n_f-0.0694n^2_f)a^3_{s, V}$ \\
$+\;(-122.191+30.871n_f-1.531n^2_f+0.011157n^3_f+\eta_f(12.222-0.7407n_f))a^4_{s, V}$
\end{tabular} \\ \hline
\begin{tabular}[c]{@{}c@{}}mMOM\\ $\xi=0$\end{tabular} & \begin{tabular}[c]{@{}c@{}} $1-a_{s, M}+(-1.0625+0.05556n_f)a^2_{s, M}+(-4.2409+2.097n_f-0.0694n^2_f)a^3_{s, M}$ \\ 
$+\;(-66.891+17.790n_f-1.091n^2_f+0.0120n^3_f+\eta_f(12.222-0.7407n_f))a^4_{s, M}$
\end{tabular} \\ \hline
\end{tabular}
\captionsetup{justification=centering}
\caption{\label{T-1} The PT series for the Adler function, $R(s)$-ratio and the coefficient function of the Bjorken polarized sum rule in QCD in the $\msbar$-, V- and the Landau mMOM schemes. The total factor $3\sum_f Q^2_f$ is omitted in the expressions for $D(Q^2)$ and $R(s)$. Here $\delta_f=(\sum_f Q_f)^2/(\sum_f Q^2_f)$ and $\eta_f=\sum_f Q_f$.}
\end{table}

The content of Table \ref{T-1} indicates that the contributions to the NS coefficient Bjorken function in all three schemes have sign-alternating structure in $n_f$. This property is also valid for the higher order corrections to the Adler function in the $\msbar$-scheme. In other cases, this feature is violated starting from the $a^3_s$ term. This fact
may be considered as the argument in favor of the 
well-known statement that the renormalon-motivated large-$\beta_0$ approximation is more pronounced 
for quantities calculated from perturbation theory in the $\msbar$-scheme in the Euclidean domain \cite{Beneke:1998ui}. Note, however, that the PT expressions for the Adler function and the coefficient function of the Bjorken polarized sum rule in the $\msbar$-scheme contain the contributions not only of the ultraviolet (UV) renormalons resulting in sign-alternating series, but also contributions of the infrared (IR) ones, which lead to the sign-constant series  (see e.g. \cite{Beneke:1998ui, Zakharov:1992bx} and references therein). The possible irregularities in low orders may be caused by cancellations between IR and UV renormalons.

One should also mention that the analytical expressions for $D(Q^2)$, $R(s)$ and $C_{Bjp}(Q^2)$ at the $\mathcal{O}(a^4_s)$ level in the mMOM-scheme with arbitrary linear covariant gauge parameter can be found in Refs.\cite{Garkusha:2018mua, Kataev:2017oqg, Molokoedov:2020kxq}. The investigation of the behavior of the $R(s)$-ratio for process $e^+e^-\rightarrow \gamma^*\rightarrow\textit{hadrons}$ 
 depending on the energy of the center of mass system and the study of its scheme dependence on the example of the three discussed renormalization schemes have been considered at $n_f=4, 5$ previously in Refs.\cite{Kataev:2015yha, Molokoedov:2020kxq}.

\section{The CBK relation in the V-scheme }

After we have received the analytic fourth-order approximations for the Adler function, the coefficient function of the Bjorken polarized sum rule and $\beta$-function in the V-scheme in the previous subsections, we are able to examine the CBK relation (\ref{GCR}) in this scheme in detail. As known the CBK relation is implemented in the class of gauge-invariant MS-like schemes at the $\mathcal{O}(a^4_s)$ level of PT at least \cite{Broadhurst:1993ru, Baikov:2010je} (but apparently in all orders  \cite{Gabadadze:1995ei, Crewther:1997ux, Braun:2003rp}). However, the following question remains open: whether this relation will be carried out in other gauge-independent schemes different  than the MS-like ones. In this section we will investigate this issue on the example of the considered gauge-invariant V-scheme.

In fact, our problem reduces to verification of the presence of the $\beta$-factorization property in the conformal symmetry breaking term $\Delta_{csb}$ in the CBK relation in the V-scheme:
\begin{equation}
\label{GCR-V}
D^{NS}_V(a_{s, V})C^{NS}_{Bjp, V}(a_{s, V})=1+\bigg(\frac{\beta^V(a_{s, V})}{a_{s, V}}\bigg)K^V(a_{s, V}).
\end{equation}

Using the V-scheme analogs of Eqs.(2.3a-2.3d) from Ref.\cite{Garkusha:2018mua}, which also follow from the formula (\ref{GCR-V}) we are testing, one can obtain that the first coefficient in expansion
\begin{equation}
\label{K-exp-V}
K^V(a_{s, V})=\sum\limits_{n\geq 1} K^V_n a^n_{s, V}
\end{equation}
coincides with its $\rm{\overline{MS}}$-scheme analog, viz
\begin{equation}
\label{K1V}
K^V_1=\bigg(-\frac{21}{8}+3\zeta_3\bigg)C_F.
\end{equation}

Similarly, utilizing the V-scheme results for $a^3_s$-corrections to the flavor NS Adler function (\ref{d3V}), the coefficient function of the Bjorken polarized sum rule (\ref{c3V}) and two-loop contribution to $\beta^V$ (\ref{beta1}), we find the second term in the expansion (\ref{K-exp-V}):
\begin{equation}
\label{K2V}
K^V_2=\bigg(\frac{397}{96}+\frac{17}{2}\zeta_3-15\zeta_5\bigg)C^2_F+\bigg(-\frac{1453}{96}+\frac{53}{4}\zeta_3\bigg)C_FC_A+\bigg(\frac{31}{8}-3\zeta_3\bigg)C_FT_Fn_f.
\end{equation}

In expression (\ref{K2V}) the analytical term, proportional to $C^2_F$-factor, is identical to its $\rm{\overline{MS}}$- \cite{Broadhurst:1993ru} and mMOM-scheme counterparts at $\xi=0$\footnote{\label{Note-1} 
As well as the mMOM-analogs in gauges $\xi=-1$ and $\xi=-3$.} \cite{Garkusha:2018mua, Kataev:2017oqg, Molokoedov:2020kxq}. However, another V-scheme abelian contribution (\ref{K2V}), containing $C_FT_Fn_f$-color structure,  
 coincides with the Landau mMOM analog \cite{Garkusha:2018mua, Kataev:2017oqg, Molokoedov:2020kxq} only, but not with $\rm{\overline{MS}}$-scheme term \cite{Broadhurst:1993ru}. The reason for this lies in the feature of determination of the mMOM-scheme \cite{vonSmekal:2009ae, Gracey:2013sca}. Indeed, the relation between couplings $a_{s, M}$ in the mMOM-scheme and $a_s$ in the $\rm{\overline{MS}}$-one  requires knowledge of the renormalization constants of the gluon and ghost fields only, but not of any vertex structures \cite{vonSmekal:2009ae}. Taking the renormalization mMOM-conditions into account, one can arrive to the following relation \cite{vonSmekal:2009ae, Gracey:2013sca, Ruijl:2017eht, Garkusha:2018mua}:
\begin{equation}
\label{asM-as}
a_{s, M}(\mu^2)=\frac{a_s(\mu^2)}{\bigg(1+\Pi_A(a_s(\mu^2), \xi(\mu^2))\bigg)\bigg(1+\Pi_c(a_s(\mu^2), \xi(\mu^2))\bigg)^2}~,
\end{equation}
where $\Pi_A$ and $\Pi_c$ are the $\rm{\overline{MS}}$-scheme gluon and ghosts self-energy functions correspondingly. They were calculated with explicit dependence on the gauge parameter $\xi$ 
at three-loop level in Ref.\cite{Chetyrkin:2000dq} and at the four-loop level in Ref.\cite{Ruijl:2017eht}. Since in the abelian limit of the $U(1)$-group all gauge-dependent terms, proportional to the
 eigenvalue $C_A$ of the Casimir operator in the adjoint representation, are nullified, and $C^i_F(T_Fn_f)^j$-contributions remain with $C_F=1$ and $T_F=1$, then the $U(1)$-analog of the formula (\ref{asM-as}) read
\begin{equation}    
\label{asM-as-QED}
a_{M}(\mu^2)=\frac{a(\mu^2)}{1+\Pi_{QED}(a(\mu^2))}=a_{{\rm{MOM}}}(\mu^2)=a_{{\rm{inv}}}(\mu^2).
\end{equation}

The l.h.s. of the formula (\ref{asM-as-QED}) matches the definition of the RG-invariant and scheme-independent invariant charge in QED, governed the  higher order corrections to the Coulomb static potential in QED (except for the light-by-light-type contributions that are appeared starting from the $\mathcal{O}(a^3)$ level and are not included in the photon vacuum polarization function in Eq.(\ref{asM-as-QED})). This fact makes the gauge-dependent mMOM- and gauge-invariant V-schemes akin in QCD. That is why the abelian $C^2_F$ and $C_FT_Fn_f$-terms in Eq.(\ref{K2V}) coincide in these two different schemes.

Using now the $a^4_s$ approximations for the NS contributions to $D(Q^2)$ (\ref{d4V}), to $C_{Bjp}(Q^2)$ (\ref{c4V}) and the three-loop expression for $\beta^V$ (\ref{beta2V}), we obtain the third term in (\ref{K-exp-V}):
\begin{eqnarray}
\label{K3V}
K^V_3&=&\bigg(\frac{2471}{768}+\frac{61}{8}\zeta_3-\frac{715}{8}\zeta_5+\frac{315}{4}\zeta_7\bigg)C^3_F \\ \nonumber
&+&\bigg(\frac{75143}{2304}+\frac{2059}{32}\zeta_3-\frac{545}{6}\zeta_5-\frac{105}{8}\zeta_7\bigg)C^2_FC_A
+
\bigg(-\frac{1273}{144}-\frac{599}{24}\zeta_3+\frac{75}{2}\zeta_5\bigg)C^2_FT_Fn_f \\ \nonumber
&+&\bigg(-\frac{71389}{576}+\frac{15235}{192}\zeta_3+\frac{2975}{48}\zeta_5-\frac{187}{8}\zeta^2_3+\frac{63}{32}\pi^2-\frac{9}{4}\pi^2\zeta_3-\frac{63}{512}\pi^4+\frac{9}{64}\pi^4\zeta_3\bigg)C_FC^2_A \\ \nonumber
&+&\bigg(\frac{40931}{576}-\frac{1771}{48}\zeta_3-\frac{125}{3}\zeta_5+\frac{17}{2}\zeta^2_3\bigg)C_FC_AT_Fn_f+\bigg(-\frac{49}{6}+\frac{7}{2}\zeta_3+5\zeta_5\bigg)C_FT^2_Fn^2_f.
\end{eqnarray}

As we have anticipated, all abelian contributions in 
(\ref{K3V}) are the same as in the mMOM-scheme in Landau gauge \cite{Garkusha:2018mua} (unlike the $\rm{\overline{MS}}$ results, where only $C^3_F$-term is equal to the V-scheme one). However, in contrast to both $\rm{\overline{MS}}$- and mMOM-scheme cases, the V-scheme $K^V_3$-coefficient contains extra $\pi^2$, $\pi^2\zeta_3$, $\pi^4$ and $\pi^4\zeta_3$-terms to $C_FC^2_A$ color structure. The other two non-abelian pieces, proportional to $C^2_FC_A$ and $C_FC_AT_Fn_f$, repeat the transcendental pattern of the $\rm{\overline{MS}}$- and Landau mMOM-scheme results for $K_3$-coefficient.

As we have already noted above, the light-by-light (l-b-l) type scattering terms with  $d^{abcd}_Fd^{abcd}_A/d_R$, $d^{abcd}_Fd^{abcd}_Fn_f/d_R$, $C_Fd^{abcd}_Fd^{abcd}_A/N_A$ and $C_Fd^{abcd}_Fd^{abcd}_Fn_f/N_A$ color factors including in $d^{NS}_{4, V}$ (\ref{d4V}) and $c^{NS}_{4, V}$ (\ref{c4V}) are canceled out exactly in the expression $d^{NS}_{4, V}+c^{NS}_{4, V}$, which is equal to:
\begin{subequations}
\begin{eqnarray}
\label{l-b-l-cancel}
(d^{NS}_{4, V}+c^{NS}_{4, V})\bigg\vert_{l-b-l}=(d^{NS}_{4}+c^{NS}_{4})\bigg\vert_{l-b-l}-a_3(d^{NS}_1+c^{NS}_1)\bigg\vert_{l-b-l}.
\end{eqnarray}

The formula (\ref{l-b-l-cancel}) follows directly from the RG-invariance of $D(Q^2)$ and $C_{Bjp}(Q^2)$ functions, and from the relation (\ref{asV-as}). One should emphasize that the discussed mutual cancellation is the consequence of the conformal symmetry. Indeed, the equality $d^{NS}_1+c^{NS}_1=0$  is the attribute of the CBK relation and arises from the non-renormalizability of the AVV triangle graph at the $\mathcal{O}(\alpha_s)$ level. This feature was confirmed by direct calculations in \cite{Jegerlehner:2005fs}. Therefore, in this order the application of the renormalization procedure does not lead to the appearance of the conformal symmetry breaking term in the CBK relation. In its turn, in the conformal invariant limit when all coefficients $\beta_k$ of the RG $\beta$-function are nullified, the sum $d^{NS}_{4}+c^{NS}_{4}$ in the $\msbar$-scheme are expressed only through terms $d_k$ and $c_k$ at $1\leq k\leq 3$ (see e.g. Eq.(2.3d) of Ref.\cite{Garkusha:2018mua}), which do not contain the l-b-l contributions. This means that the equality $(d^{NS}_{4}+c^{NS}_{4})\vert_{l-b-l}=0$ is the consequence of the conformal symmetry \cite{Kataev:2010du}. Based on these arguments, we conclude that l.h.s. of Eq.(\ref{l-b-l-cancel}) is identically equal to zero:
\begin{equation}
\label{sum l-b-l 0}
(d^{NS}_{4, V}+c^{NS}_{4, V})\bigg\vert_{l-b-l}=0.
\end{equation}
\end{subequations}

This feature is extremely important for factorization of the $\beta$-function in the CBK relation at the $\mathcal{O}(\alpha^4_s)$ level. In a full analogy, one can show that the terms, originating from  expression (\ref{a3}) for $a_3$-coefficient and proportional to $\pi^6$, $s_6$, $\pi^4\log 2$, $\pi^4\log^2 2$, $\pi^2\alpha_4$, $\pi^2\log 2$ and $\pi^2\zeta_3\log 2$-numbers, are 
mutually canceled out in the sum $d^{NS}_{4, V}+c^{NS}_{4, V}$.  This fact is the consequence of the conformal symmetry as well\footnote{The interesting consequences of the violation of the conformal symmetry are briefly discussed in Appendix B.}. 

So, we have demonstrated that the factorization of $\beta^V$-function in the CBK relation in the V-scheme is performed in the fourth order of PT  at least. The natural question arises: will this factorization property be also true in other  gauge-invariant renormalization schemes other than the V- or MS-like ones? We give response to this issue in the next section.

\section{The CBK relation in different gauge-invariant schemes}

In this section we continue to develop the ideas proposed in Ref.\cite{Garkusha:2018mua}. There were investigated the requirements for the gauge-dependent schemes that would ensure fulfillment of the CBK relation. It turns out that if the CBK relation in QCD is valid in the $\msbar$-scheme in all orders of PT (the grounds to trust this assumption are given in Refs.\cite{Gabadadze:1995ei, Crewther:1997ux, Braun:2003rp}), then it will be true for a wide class of the MOM-like schemes with linear covariant Landau gauge in all orders as well. Let us now extend the ideas
of Ref.\cite{Garkusha:2018mua} applied to a class of the gauge-invariant renormalization schemes.

We will conduct our study for the particular case of the V-scheme. Without limiting generality, {\textit{it is possible to understand by the V-scheme any other gauge-invariant scheme}} whose coupling constant is related to $a_s$ in the $\msbar$-scheme by the relation \`a-la (\ref{asV-as}) with some coefficients $a_k$. 

Taking into account the RG-invariance of functions $D^{NS}(a_s)$ and $C^{NS}_{Bjp}(a_s)$ and using the explicit form of the conformal symmetry breaking term in the CBK relation, we gain the following equality
\begin{equation}
\label{K-MS-V}
\frac{\beta(a_s)}{a_s}K(a_s)=\frac{\beta^V(a_{s, V}(a_s))}{a_{s, V}(a_s)}K^V(a_{s, V}(a_s)),
\end{equation}
linking the polynomials $K(a_s)$ and $K^V(a_{s, V})$ in the $\rm{\overline{MS}}$- and V- ({\textit{arbitrary gauge-invariant}}) scheme correspondingly. 
Utilizing the scheme independence of the first two coefficients of the RG $\beta$-function (\ref{beta0}-\ref{beta1}) within a class of the gauge-invariant schemes and substituting Eq.(\ref{asV-as}) in (\ref{K-MS-V}), we get the following relationships between the V- and $\rm{\overline{MS}}$-scheme coefficients of the polynomial $K(a_s)$:
\begin{eqnarray}
\label{K-V-MS-rel-1}
K^V_1&=&K_1, \\
\label{K-V-MS-rel-2}
K^V_2&=&K_2-2a_1K_1.
\end{eqnarray}

The $\msbar$ coefficients $K_1$ and $K_2$ were first calculated in \cite{Broadhurst:1993ru}. Accommodating their explicit form, we find that formulas (\ref{K-V-MS-rel-1}) and (\ref{K-V-MS-rel-2}) are in full agreement with results of the direct calculation (\ref{K1V}) and (\ref{K2V}). The expression (\ref{K-V-MS-rel-2}) is analogous to Eq.(4.3) of Ref.\cite{Garkusha:2018mua} that presents the relation between $K_2$-coefficients in the mMOM and $\rm{\overline{MS}}$-scheme. However, unlike this attitude, the equation (\ref{K-V-MS-rel-2}) does not contain terms with $1/\beta_0$-factor. This fact is a consequence of scheme independence of  two-loop coefficient $\beta_1$ in a class of the gauge-invariant renormalization schemes like the V-scheme. Remind that in the gauge-dependent schemes, such as mMOM,  the two-loop coefficient of the $\beta$-function already depends on gauge. Despite this fact, it was found out in Ref.\cite{Garkusha:2018mua} that the CBK relation holds in the mMOM-scheme in the $\mathcal{O}(a^3_s)$ approximation  for three values of the gauge parameter only, namely for $\xi=0, -1, -3$.  The mentioned $1/\beta_0$-term disappears at these values of $\xi$.

Carrying out similar steps in the next order PT, we obtain
\begin{equation}
\label{interm-K-V-MS}
K^V_3=K_3-3a_1K_2+\bigg(\frac{\beta_2-\beta^V_2-a_1\beta_1}{\beta_0}+5a^2_1-2a_2\bigg)K_1.
\end{equation}

Taking now into account Eq.(\ref{b2}), we derive the  simplified form of Eq.(\ref{interm-K-V-MS}) without $1/\beta_0$-factor:
\begin{equation}
\label{K-V-MS-rel-3}
K^V_3=K_3-3a_1K_2+(6a^2_1-3a_2)K_1.
\end{equation}

The coefficient $K_3$ is known from results of work \cite{Baikov:2010je}. The application of the formula (\ref{K-V-MS-rel-3}) reproduces the expression  (\ref{K3V}). Note also that the final structure of Eq.(\ref{K-V-MS-rel-3}) is much simpler than the analogous one, presented in Eq.(4.7) of Ref.\cite{Garkusha:2018mua} for the mMOM-scheme with arbitrary gauge. As was shown in this quoted work,
at the $\mathcal{O}(a^4_s)$ level the CBK relation remains valid in the mMOM-scheme (and other QCD MOM-like schemes) only for the Landau gauge $\xi=0$.

Further, proceeding in a similar way and employing Eqs.(\ref{b2}-\ref{b4}), we straightforwardly get 
\begin{eqnarray}
\label{K-V-MS-rel-4}
K^V_4&=&K_4-4a_1K_3+(10a^2_1-4a_2)K_2+(20a_1a_2-20a^3_1-4a_3)K_1, \\
\label{K-V-MS-rel-5}
K^V_5&=&K_5-5a_1K_4+(15a^2_1-5a_2)K_3+(30a_1a_2-35a^3_1-5a_3)K_2 \\ \nonumber
&+&(30a_1a_3+15a^2_2-105a^2_1a_2+70a^4_1-5a_4)K_1.
\end{eqnarray}
where $K_4$ and $K_5$ are still unknown $\rm{\overline{MS}}$-scheme coefficients of Eq.(\ref{GCR}), $a_4$ is yet completely unknown four-loop correction to the static potential in QCD (\ref{V}). 

There are no obstacles to obtain similar representations of $K^V_n$-terms for any order of PT. Thus, we are convinced that if the CBK relation in QCD is valid in the $\msbar$-scheme in all orders of PT, then it will be also true for arbitrary gauge-invariant renormalization scheme with ``non-exotic'' coefficients $a_k$ included in the ratio between couplings in this considered scheme and in the $\msbar$-scheme (see analog of Eq.(\ref{asV-as})). Under ``non-exotic'' coefficients we imply such ones that are the polynomials in $n_f$ with coefficients which are algebraic or transcendental numbers. For example, in the gauge-invariant 't Hooft scheme \cite{'t Hooft 1, 't Hooft 2} the coefficients $a_k$ are not polynomials in $n_f$, they contain terms proportional to $1/\beta_0$-factors. The feature of this scheme lies in the fact that its $\beta$-function contains two nonzero scheme-independent PT coefficients only and the rest are assumed to be zero by finite renormalization of charge. As was shown in Ref.\cite{Garkusha:2011xb}, the transition from the $\msbar$-scheme to the 't Hooft scheme spoils the property of factorization of the $\beta$-function in the CBK relation.

\section{The QED case}

Let us now consider the QED case with $N$ charged leptons. The transition to the abelian $U(1)$ gauge group is performed by replacing $C_A=0$, $C_F=1$, $T_F=1$, $d^{abcd}_A=0$, $d^{abcd}_{F}=1$, $d^{abc}=1$, $N_A=1$, $d_R=1$, $n_f=N$.

Using the analytical expression for the four-loop approximation of $\beta$-function in the V-scheme 
in the case of the generic simple gauge group (\ref{beta0}-\ref{beta3V}) and taking into account the transition discussed above, one can obtain the  QED analog of the $\beta$-function in the V-scheme:
\begin{eqnarray}
\label{QEDV}
\beta^V_{QED}(a_V)&=&\frac{1}{3}Na_V^2+\frac{1}{4}Na_V^3
+\bigg(-\frac{1}{32} N + \bigg(\frac{1}{3}\zeta_3-\frac{23}{72}\bigg)N^2\bigg)a_V^4 \\ \nonumber
&+&\bigg(-\frac{23}{128} N+\bigg(\frac{13}{32} +\frac{2}{3}\zeta_3-\frac{5}{3}\zeta_5+\frac{2}{3}\mathcal{C}\bigg)N^2
+\bigg(\frac{1}{2}-\frac{1}{3}\zeta_3\bigg)N^3\bigg)a_V^5,
\end{eqnarray}
where $a_V=\alpha_V/\pi$. The introduced constant $\mathcal{C}$ arises naturally in the definition of the static Coulomb potential through the QED invariant charge:
\begin{equation}
\label{potential-QED-inv}
V_{QED}(\vec{q}^{\;2})=-\frac{4\pi}{\vec{q}^{\;2}}\frac{\alpha(\mu^2)}{1+\Pi_{QED}(\vec{q}^{\;2}/\mu^2, \; \alpha)}\bigg(1+N\cdot \mathcal{C} \bigg(\frac{\alpha}{\pi}\bigg)^3+\dots\bigg).
\end{equation}

Here $\mathcal{C}$ is the correction associated with the appearance of the light-by-light type scattering diagrams to the static potential \cite{Lee:2016cgz}, which do not occur in the photon vacuum polarization function $\Pi_{QED}$ at this level \cite{Gorishnii:1991hw}:
\begin{align}
\label{math-C}
\mathcal{C}=&~\frac{5}{96}\pi^6
-\pi^4\bigg(\frac{23}{24}-\frac{\log 2}{6}+\frac{\log^2 2}{2}\bigg)+\pi^2\bigg(\frac{79}{36}-\frac{61}{12}\zeta_3+\log 2+
\frac{21}{2}\zeta_3\log 2\bigg) 
\approx -0.888062.
\end{align}

The numerical effect of the term $2\mathcal{C}N^2a^5_V/3$ in (\ref{QEDV}) is not negligible compared to the contribution of the remaining part, proportional to $N^2a^5_V$, but on the contrary, it even dominates it.

The expression (\ref{QEDV}) should be compared with the QED result for $\beta$-function in the MOM-scheme identical to the Gell-Mann--Low $\Psi$-function  \cite{Gorishnii:1990kd}:
\begin{eqnarray}
\label{QEDMOM} 
\hspace{-0.8cm}
\beta^{{\rm{MOM}}}_{QED}(a_{{\rm{MOM}}})=\Psi(a_{{\rm{MOM}}})  &=&\frac{1}{3}Na_{{\rm{MOM}}}^2+\frac{1}{4} Na_{{\rm{MOM}}}^3+
\bigg(-\frac{1}{32} N + \bigg(\frac{1}{3}\zeta_3-\frac{23}{72}\bigg)N^2\bigg)a_{{\rm{MOM}}}^4
\\ \nonumber
&+&\bigg(-\frac{23}{128} N+\bigg(\frac{13}{32} +\frac{2}{3}\zeta_3-\frac{5}{3}\zeta_5\bigg)N^2+ 
\bigg(\frac{1}{2} -\frac{1}{3}\zeta_3\bigg)N^3\bigg)a_{{\rm{MOM}}}^5,
\end{eqnarray}
where $a_{\rm{MOM}}=\alpha_{\rm{MOM}}/\pi$ coincides with the QED invariant charge (\ref{asM-as-QED}). Note that the expression (\ref{QEDMOM}) may be also obtained  e.g. as a result of the transition to the case of the $U(1)$ group for the $\beta$-function computed initially in the mMOM-scheme in the generic simple gauge group \cite{Gracey:2013sca, Ruijl:2017eht}. This fact is explained and directly follows from the formulas  (\ref{asM-as}) and (\ref{asM-as-QED}).

One can see that at the three-loop level $\beta^V_{QED}$ (\ref{QEDV}) completely coincides in form with the Gell-Mann--Low $\Psi$-function (\ref{QEDMOM}). The difference between them starts to manifest itself only at the four-loop level due to the additional term $2\mathcal{C}N^2a^5_V/3$, related to the light-by-light type scattering effect in the perturbative expression for the static Coulomb potential (\ref{potential-QED-inv}):
\begin{equation}
\label{diff-V-Psi}
\beta^V_{3, QED}=\Psi_3+\frac{2}{3}\mathcal{C}N^2.
\end{equation}

The obtained result may be presented in the following compact form:
\begin{equation}
\label{a-V-MOM}
a_V=a_{{\rm{MOM}}}+\mathcal{C}Na^4_{{\rm{MOM}}}+\mathcal{O}(a^5_{{\rm{MOM}}}).
\end{equation}

The arguments, given in this Section and in Sec.4, enables us to conclude that in QCD the V-scheme possesses many similar properties as the MOM-like schemes in the Landau gauge. Since difference between couplings $a_V$ and $a_{{\rm{MOM}}}$ in QED (\ref{a-V-MOM}) starts to reveal itself from the $\alpha^4$-term only, then, with some reservations, the V-scheme in QCD may be interpreted as the gauge-independent scheme in which one can construct an analog of the gauge-invariant charge, namely the gauge-invariant combinations of the related Green functions. Remind, it is impossible to introduce this concept within the gauge-dependent MOM-like schemes in QCD. For details see e.g. \cite{Kataev:2015yha}.

Setting $N=1$ in Eqs.(\ref{QEDV}) and (\ref{QEDMOM}), we arrive to their numerical form:
\begin{eqnarray}
\label{numv}
\beta^{V}_{QED}(a_V)&=&0.3333a_{V}^2+0.25a_{V}^3+0.0499a_{V}^4-
1.19301a_{V}^5, \\
\label{numom}
\Psi(a_{\rm{MOM}})&=&0.3333a_{\rm{MOM}}^2+0.25a_{\rm{MOM}}^3+0.0499a_{\rm{MOM}}^4-
0.60096a_{\rm{MOM}}^5.
\end{eqnarray}

One can observe that even at $N=1$  
the numerical effect of the light-by-light scattering 
contribution, which is typical for the V-scheme,
is rather sizable and is almost equal to twice four-loop correction to the $\Psi$-function.

Let us turn to the consideration of relations between higher order corrections to $\beta^V_{QED}$ and $\Psi$-function. The dependence of these RG-functions on the number $N$ of the charged leptons is described via the following decompositions:
\begin{eqnarray} 
\label{betaVN}
\beta^{V}_{QED}(a_{V})&=&
\beta^{V(1)}_{QED, 0}Na_{V}^2+\sum\limits_{i\geq 1}\sum\limits_{k=1}^i\beta^{{V}(k)}_{QED, i}N^ka_{V}^{i+2}, \\
\label{PsiVN}
\Psi(a_{\rm{MOM}})&=&\Psi^{(1)}_0Na_{\rm{MOM}}^2+
\sum\limits_{i\geq 1}\sum\limits_{k=1}^i\Psi^{(k)}_iN^ka_{\rm{MOM}}^{i+2}.
\end{eqnarray}

As we have already seen, the coefficients $\beta^{{V}(k)}_{QED, i}$ and $\Psi^{(k)}_i$ will differ only by corrections $\Delta\beta_{QED, i}^{{V}(k)}$ associated to the light-by-light scattering-type effects in the static potential: 
 \begin{equation}
\label{exrtaN}
\beta_{QED, i}^{{V(k)}}=\Psi_{i}^{(k)}+\Delta\beta_{QED, i}^{{V(k)}}.
\end{equation}

This fact directly follows from the definition of the coupling constant $a^V$ in the V-scheme (QED-analog of Eq.(\ref{V-schemedef}) and (\ref{potential-QED-inv})) and from relation (\ref{asM-as-QED}). It is clear that the extra term $\Delta\beta_{QED, i}^{{V(k)}}$ will appear only for indexes $\{i,k\}=\{i\geq 3,~ 2\leq k\leq i-1\}$. In cases when $\{i,k\}=
\{i\geq 3,~ k=1 \;or\; k=i\}$, the coefficients of $\beta^V$ and $\Psi$-functions coincide. Indeed, we have already observed that at the four-loop level
 \begin{equation}
\beta_{QED, 3}^{{V(1)}}=\Psi_3^{(1)}=
-\frac{23}{128}, ~~~~~
\beta_{QED, 3}^{{V(3)}}=\Psi_3^{(3)}= 
\frac{1}{2}-\frac{1}{3}\zeta_3.
\end{equation}

The RG $\beta$-function in the MOM-scheme (the Gell-Mann--Low $\Psi$-function) was calculated in the fifth-loop approximation in QED in \cite{Baikov:2012zm} for arbitrary $N$ (and in the $\msbar$-scheme as well):
\begin{eqnarray}
\label{Psi4}
\Psi_4&=&\bigg(\frac{4157}{6144}+\frac{1}{8}\zeta_3\bigg)N+\bigg(-\frac{251}{256}-\frac{23}{24}\zeta_3-\frac{45}{8}\zeta_5+\frac{35}{4}\zeta_7\bigg)N^2 \\ \nonumber
&+&\bigg(-\frac{3383}{3456}-\frac{205}{72}\zeta_3+\frac{5}{2}\zeta_5+\zeta^2_3\bigg)N^3+\bigg(-\frac{67}{72}+\frac{7}{18}\zeta_3+\frac{5}{9}\zeta_5\bigg)N^4.
\end{eqnarray}

For instance, this result may be obtained as the $U(1)$-limit of the $\beta$-function computed at the five-loop level in the mMOM-scheme with arbitrary gauge parameter in the case of the generic simple gauge group in Ref.\cite{Ruijl:2017eht}.

Using now the formula (\ref{b4}) and the expression for $\beta_4$ in the $\msbar$-scheme \cite{Baikov:2012zm}, we find the five-loop coefficient of $\beta^V_{QED}$-function:
\begin{align}
\hspace{-0.5cm}
\label{b4VQED-a}
&\beta^V_{QED,4}=\bigg(\frac{4157}{6144}+\frac{1}{8}\zeta_3\bigg)N+\bigg(-\frac{49}{48}-\frac{53}{96}\zeta_3+\frac{65}{32}\zeta_5+\frac{1}{4}\mathcal{C}+a^{(1)}_4\bigg)N^2 \\ \nonumber
&+\bigg(-\frac{4255}{4608}+\frac{1013}{144}\zeta_3-\frac{13}{96}\zeta_4-\frac{215}{36}\zeta_5-\frac{5}{3}\zeta^2_3+\frac{20}{9}\mathcal{C}+a^{(2)}_4\bigg)N^3+\bigg(\frac{118907}{31104}-\frac{71}{24}\zeta_3+a^{(3)}_4\bigg)N^4,
\end{align}
where the constant $\mathcal{C}$ has been defined above and, in analogy with Eq.(\ref{a3}), we have utilized the decomposition of the four-loop correction $a_4$ to the static Coulomb potential in QED in powers of $N$:
\begin{equation}
\label{a4-decomp}
a_4=\bigg(\frac{5}{9}\bigg)^4 N^3+a^{(3)}_4N^3+a^{(2)}_4N^2+a^{(1)}_4N.
\end{equation}

One should note that the term $-13\zeta_4/96$ in the $N^3$-coefficient of $\beta^V_{QED, 4}$ (\ref{b4VQED-a}) is not related to the effects of the light-by-light scattering but arises from the calculations of $\beta_4$ in the $\msbar$-scheme (see \cite{Baikov:2012zm} and   \cite{Herzog:2017ohr, Luthe:2017ttg}).

As we have expected, the terms linear in $N$ are the same in Eqs.(\ref{Psi4}) and (\ref{b4VQED-a}):
\begin{equation}
\label{b4V-Psi4}
\beta_{QED, 4}^{{V(1)}}=\Psi_4^{(1)}=\frac{4157}{6144}+\frac{1}{8}\zeta_3.
\end{equation}

It was explained in Ref.\cite{Kataev:2013vua} that the scheme-independence of these linear terms in the massless QED is the consequence of the conformal symmetry.

Since coefficients $\beta_{QED, 4}^{{V(4)}}$ and $\Psi_4^{(4)}$ should also be the same, then we can fix the contribution $a^{(3)}_4$ from matching Eqs.(\ref{Psi4}) and (\ref{b4VQED-a}):
\begin{eqnarray}
\label{a43}
a^{(3)}_4&=&-\frac{147851}{31104}+\frac{241}{72}\zeta_3+\frac{5}{9}\zeta_5, \\ 
\beta_{QED, 4}^{{V(4)}}&=&\Psi_4^{(4)}=-\frac{67}{72}+\frac{7}{18}\zeta_3+\frac{5}{9}\zeta_5.
\end{eqnarray}

The four-loop expressions for $\beta_{QED, 4}^{{V(2)}}$ and $\beta_{QED, 4}^{{V(3)}}$ will contain the contributions, related to the light-by-light scattering-type effects in the static potential. They appear and are mixed both from the constant $\mathcal{C}$, occurring at the three-loop level, and from the fourth-order corrections $a^{(1)}_4$ and $a^{(2)}_4$ (\ref{b4VQED-a}). Based on results of Ref.\cite{Lee:2016cgz}, one can conclude that the contributions of these effects are separated from other ones by transcendent constants proportional to   even powers of the $\pi$-number (see Eq.\ref{math-C}). Without these still unknown terms, the corrections $a^{(1)}_4$ and $a^{(2)}_4$ read:
\begin{eqnarray}
\label{a41}
a^{(1)}_4\bigg\vert_{no \;l-b-l}&=&\frac{31}{768}-\frac{13}{32}\zeta_3-\frac{245}{32}\zeta_5+\frac{35}{4}\zeta_7, \\ 
\label{a42}
a^{(2)}_4\bigg\vert_{no \;l-b-l}&=&-\frac{767}{13824}-\frac{1423}{144}\zeta_3+\frac{13}{96}\zeta_4+\frac{305}{36}\zeta_5+\frac{8}{3}\zeta^2_3.
\end{eqnarray}

These expressions directly follow from equating $\Psi^{(2)}_4$ to $\beta_{QED, 4}^{{V(2)}}$ and $\Psi^{(3)}_4$ to $\beta_{QED, 4}^{{V(3)}}$ in approximation when the light-by-light scattering effects in the static potential are discarded. In its turn, the following relations are valid:
\begin{eqnarray}
\label{lbl-Psi-V-2}
\beta_{QED, 4}^{{V(2)}}&=&\Psi_{4}^{(2)}+\Delta\beta_{QED, 4}^{{V(2)}}, ~~~~ \Delta\beta_{QED, 4}^{{V(2)}}=a^{(1)}_4\bigg\vert_{l-b-l}+\frac{1}{4}\mathcal{C},     \\
\label{lbl-Psi-V-3}
\beta_{QED, 4}^{{V(3)}}&=&\Psi_{4}^{(3)}+\Delta\beta_{QED, 4}^{{V(3)}}, ~~~~ \Delta\beta_{QED, 4}^{{V(3)}}=a^{(2)}_4\bigg\vert_{l-b-l}+\frac{20}{9}\mathcal{C}.
\end{eqnarray}

Formulas (\ref{a43}), (\ref{a41}) and (\ref{a42})
are generalized without significant obstacles to the case of the generic simple gauge group and then look more clearly:
\begin{eqnarray}
\label{a34-SU}
a^{(3)}_4\bigg\vert_{abelian}&=&\bigg(-\frac{147851}{31104}+\frac{241}{72}\zeta_3+\frac{5}{9}\zeta_5\bigg)C_FT^3_F, \\
\label{a24-SU}
a^{(2)}_4\bigg\vert_{abelian, \; no \;l-b-l}&=&\bigg(\frac{13025}{13824} - \frac{403}{36}\zeta_3-\frac{11}{96}\zeta_4+\frac{175}{18}\zeta_5+2\zeta^2_3\bigg)C^2_FT^2_F  \\ \nonumber
&+&\bigg(-\frac{431}{432}+\frac{21}{16}\zeta_3+\frac{1}{4}\zeta_4-\frac{5}{4}\zeta_5+\frac{2}{3}\zeta^2_3\bigg)\frac{d^{abcd}_Fd^{abcd}_F}{N_A},   \\
\label{a14-SU}
a^{(1)}_4\bigg\vert_{abelian, \; no \;l-b-l}&=&\bigg(\frac{31}{768}-\frac{13}{32}\zeta_3-\frac{245}{32}\zeta_5+\frac{35}{4}\zeta_7\bigg)C^3_FT_F.
\end{eqnarray}

The $d^{abcd}_Fd^{abcd}_F$-contribution to $a^{(2)}_4$ (\ref{a24-SU}) originates from $d^{abcd}_Fd^{abcd}_F$-ones to coefficients $\beta_3$ and $\beta_4$. This fact may be directly observed from Eq.(\ref{b4}), where the abelian terms, proportional to $C_F$ and $d^{abcd}_Fd^{abcd}_F$, may be fixed  from consideration $\beta^{{\rm{mMOM}}}_4$ \cite{Ruijl:2017eht} (whose the abelian contributions in the Landau gauge are equal to those in $\beta^V_4$ without taking into account the light-by-light scattering-type corrections to the static potential) and from analytical results for $\beta_3$ and $\beta_4$ \cite{Herzog:2017ohr, Luthe:2017ttg}. One should emphasize that expressions (\ref{a34-SU}-\ref{a14-SU}) are in full agreement with the analogous results presented in Eq.(14.4) of Ref.\cite{Grozin:2022umo}.

\section{Conclusion}

In this work we obtain the explicit analytical form of the RG $\beta$-function in the gauge-invariant $V$-scheme at the four-loop level in the case of the generic simple gauge group. Using the renormalization invariance of the Adler function for process of  $e^+e^-\rightarrow \gamma^*\rightarrow {\textit{hadrons}}$, $R_{e^+e^-}(s)$-ratio and the coefficient function of the Bjorken polarized sum rule of deep-inelastic scattering of the polarized charged leptons on nucleons, we get their PT expressions in the V-scheme up to $\alpha^4_s$-corrections as well. The comparison of the derived V-scheme results with the $\msbar$- and mMOM-counterparts in the Landau gauge is performed. In the cases of the Adler function and $R_{e^+e^-}(s)$-ratio in the V- and mMOM-schemes the nonregular behavior of the perturbative corrections in their decomposition in powers of $n_f$ is observed in higher orders. Taking into account the obtained V-scheme results, we demonstrate explicitly that the  CBK relation remains valid in this effective scheme at the $\mathcal{O}(\alpha^4_s)$ level. Further, we prove our hypothesis that factorization of the RG $\beta$-function in the conformal symmetry breaking term of the CBK relation will be true in any gauge-invariant scheme at least in the fourth order of PT. The  chosen gauge-invariant scheme should only lead to the ``non-exotic'' coefficients in the relationship between couplings defined in the $\msbar$-scheme and in the considered one, i.e. these coefficients should be polynomials in $n_f$. Moreover, it turns out that if the CBK relation in QCD is valid in the $\msbar$-scheme in all orders of PT, then it will be  true for the discussed gauge-invariant class of the renormalization schemes in all orders as well. We show that in QED  the coefficients of the $\beta$-function in the V-scheme coincide with the analogous ones in the MOM-scheme at the three-loop level. Starting from the fourth order of PT their $N^2$-coefficients begin to differ on correction associated with the manifestation of the effects of the light-by-light scattering in the static potential. The rest terms proportional to $N$ and $N^3$ stay the same. In even higher orders, this tendency will continue, i.e. two $N$-dependent terms in the coefficients of the perturbative expansions of the $\beta^V_{QED}$ and $\Psi$-functions will always coincide, and the remaining ones will differ by the correction related to the light-by-light scattering in the static potential. Based on these findings, we predict several contributions to the four-loop correction to the static potential in the case of the generic simple gauge group.

\section*{Acknowledgments}

The work of VSM was supported by the Russian Science Foundation, agreement no. 21-71-30003.

\section*{Appendix A}

Let us to consider the question related to the integral representation of the multiple zeta values. In general, these functions are defined as
\begin{equation}
\zeta_{m_1, \dots, m_k}=\sum_{i_1=1}^{\infty}\sum_{i_2=1}^{i_1-1}\dots
\sum_{i_k=1}^{i_{k-1}-1}\prod\limits_{j=1}^{k}\frac{{\rm{sign}}(m_j)^{i_j}}{i_{j}^{|m_j|}}.
\end{equation}

They were studied in detail in the number of works on the subject (see e.g. Refs.\cite{Blumlein:2009cf, Blumlein:2009fz}, \cite{Anzai:2012xw}, \cite{Lee:2015eva}). We use the Hurwitz--Lerch zeta function $\Phi(z, s, q)$
\begin{equation}
\Phi(z, s, q)=\sum\limits_{k=0}^{\infty} \frac{z^k}{(k+q)^s}
\end{equation}
and its integral representation 
\begin{equation}
\Phi(z, s, q)=\frac{1}{\Gamma(s)}\int\limits_0^1\frac{x^{q-1}(-\log x)^{s-1}}{1-zx}dx,
\end{equation}
which is valid for ${\rm{Re}}(q)>0$, ${\rm{Re}}(s)>0$ and 
$z\in [-1; 1)$ or ${\rm{Re}}(s)>1$ and $z=1$. 

Then, for the constant $\zeta_{-5,-1}$ with   transcendence of weight 6, appearing in the process of calculation of the three-loop correction to the static potential \cite{Lee:2016cgz}, it is possible to write \cite{Molokoedov:2020kxq}:
\begin{eqnarray}
\zeta_{-5,-1}&=&\sum\limits_{k=1}^{\infty}\frac{(-1)^k}{k^5}\sum\limits_{i=1}^{k-1}\frac{(-1)^i}{i}=\frac{15}{16}\zeta_5\log 2-\sum\limits_{k=1}^{\infty}\frac{\Phi(-1,1,k)}{k^5} \\ \nonumber
&=&\frac{15}{16}\zeta_5\log 2-\int\limits_0^1\frac{dx}{x(x+1)}\sum\limits_{k=1}^{\infty}\frac{x^k}{k^5}=\frac{15}{16}\zeta_5\log 2-\zeta_6+\int\limits_0^1dx \frac{{\rm{Li}}_5(x)}{x+1}.
\end{eqnarray}

Therefore, the constant $s_6=\zeta_6+\zeta_{-5,-1}$ may be presented in the following form \cite{Blumlein:2009fz}:
\begin{equation}
s_6=\frac{15}{16}\zeta_5\log 2+\int\limits_0^1dx \frac{{\rm{Li}}_5(x)}{x+1}\approx 0.9874414.
\end{equation}

Similarly, one can obtain the integral representations for the following multiple zeta values with concrete arguments arising in the intermediate calculations in the work \cite{Lee:2016cgz}:
\begin{eqnarray}
\zeta_{5, 2}&=&\zeta_5\zeta_2-\zeta_7+\int\limits_0^1 dx \frac{{\rm{Li}_5}(x)\log x}{1-x}\approx 0.0385751, \\
\zeta_{-5, 2}&=&-\frac{15}{16}\zeta_5\zeta_2+\frac{63}{64}\zeta_7+\int\limits_0^1 dx \frac{{\rm{Li}_5}(-x)\log x}{1-x}\approx 0.0271089.
\end{eqnarray}

For instance, the function $\zeta_{5,3}$ occurs during the computation of the $\msbar$-scheme $\beta$-function of the $O(N)$-symmetric $\phi^4$ theory in the six-loop approximation \cite{Kompaniets:2017yct} (in notations of this quoted paper $\zeta_{3,5}$):
 \begin{equation}
\zeta_{5, 3}=\zeta_3 \zeta_5-\zeta_8-\frac{1}{2}\int\limits_0^1 dx \frac{{\rm{Li}_5}(x)\log^2(x)}{1-x}\approx 0.0377077. 
 \end{equation}

\section*{Appendix B}

It is interesting to note some common features of the CBK relation and the action sum rule \cite{Michael:1986yi, Rothe:1995hu, Dosch:1995fz, Shoshi:2002rd, Chernodub:2010sq} (in lattice QCD also known as the Michael sum rule).  Indeed,  both of them contain a conformal anomaly term, reflecting the effect of  violation of the conformal symmetry. However, the second relation 
may be directly used in the nonperturbative region as well.

Remind that the conformal anomaly in the trace of the energy-momentum tensor of the massless $SU(N_c)$ gauge theory in the Euclidean domain has the following form \cite{Nielsen:1977sy, Adler:1976zt, Collins:1976yq}:
\begin{equation}
\label{trace}
T_{\mu\mu}(x)=\frac{\beta(a_s)}{2a_s}F^a_{\mu\nu}(x)F^a_{\mu\nu}(x)=2\frac{\beta(a_s)}{a_s}\mathscr{L}(x),
\end{equation}
where $\mathscr{L}(x)$ is the gluonic gauge part of the {\it{Euclidean}} Lagrangian density of the $SU(N_c)$ theory, expressed trough {\it{Euclidean}} chromoelectric and chromomagnetic fields
\begin{equation}
\label{chromo}
\mathscr{L}(x)=\frac{1}{4}F^a_{\mu\nu}(x)F^a_{\mu\nu}(x)=\frac{1}{2}(\vec{E}(x)^2+\vec{B}(x)^2).
\end{equation}

Note that owing to a change in a metric signature, the square of the Euclidean electric field has an opposite sing to its Minkowskian counterpart, while signs of the squares of the Euclidean and Minkowskian magnetic fields coincide. The action sum rule relates the certain combination of the static potential to the Euclidean chromoelectric and chromomagnetic condensates and $\beta$-function \cite{Michael:1986yi, Rothe:1995hu, Dosch:1995fz, Shoshi:2002rd, Chernodub:2010sq}:
\begin{equation}
\label{ASR}
\tilde{V}(r)+r\frac{\partial\tilde{V}(r)}{\partial r}=\frac{\beta(a_s)}{a_s}\langle\int d^3x(\vec{E}(x)^2+\vec{B}(x)^2)\rangle_r,
\end{equation}
where $\tilde{V}(r)$ is the static potential in the coordinate space including the confining and nonconfining components and $\langle \dots \rangle_r$ is the vacuum expectation value in the presence of a static quark-antiquark pair spaced apart from each other at a distance $r$ excluding the analogous contribution without these field sources. 

It would be interesting to study the possible relationship of the action sum rule and the CBK relation based on the first principles of quantum field theory.

\begin{flushleft}

\end{flushleft} 

\begin{thebibliography}{99}

\bibitem{Bali:2000gf}
G.~S.~Bali,
Phys. Rept. \textbf{343}, 1-136 (2001)
[arXiv:hep-ph/0001312 [hep-ph]].

\bibitem{Karbstein:2018mzo}
F.~Karbstein, M.~Wagner and M.~Weber,
Phys. Rev. D \textbf{98}, no.11, 114506 (2018)
[arXiv:1804.10909 [hep-ph]].

\bibitem{BornyakovVG:2023rci}
V.~G.~Bornyakov and I.~E.~Kudrov,
[arXiv:2301.03076 [hep-lat]].
 
\bibitem{Brambilla:1999qa}
N.~Brambilla, A.~Pineda, J.~Soto and A.~Vairo,
Phys. Rev. D \textbf{60}, 091502 (1999)
[arXiv:hep-ph/9903355 [hep-ph]].
 
\bibitem{Brambilla:2004jw}
  N.~Brambilla, A.~Pineda, J.~Soto and A.~Vairo,
Rev.\ Mod.\ Phys.\  {\bf 77} (2005) 1423
 [hep-ph/0410047].
 
\bibitem{Kniehl:2002br}
B.~A.~Kniehl, A.~A.~Penin, V.~A.~Smirnov and M.~Steinhauser,
Nucl. Phys. B \textbf{635}, 357-383 (2002)
[arXiv:hep-ph/0203166 [hep-ph]].

\bibitem{Fischler:1977yf} 
  W.~Fischler,
  Nucl.\ Phys.\ B {\bf 129} (1977) 157.

\bibitem{Billoire:1979ih} 
  A.~Billoire,
  Phys.\ Lett.\ B {\bf 92}, 343 (1980) 343.
 
\bibitem{Peter:1996ig}
M.~Peter,
Phys. Rev. Lett. \textbf{78}, 602-605 (1997)
[arXiv:hep-ph/9610209 [hep-ph]].
 
\bibitem{Schroder:1998vy}
  Y.~Schroder,
  Phys.\ Lett.\ B {\bf 447}  (1999) 321. 
  [hep-ph/9812205].

\bibitem{Gorishnii:1991hw} 
  S.~G.~Gorishnii, A.~L.~Kataev and S.~A.~Larin,
  Phys.\ Lett.\ B {\bf 273}, 141 (1991)
  Erratum: [Phys.\ Lett.\ B {\bf 275}, 512 (1992)]
  Erratum: [Phys.\ Lett.\ B {\bf 341}, 448 (1995)].

\bibitem{Smirnov:2008pn}
A.~V.~Smirnov, V.~A.~Smirnov and M.~Steinhauser,
Phys. Lett. B \textbf{668}, 293-298 (2008)
[arXiv:0809.1927 [hep-ph]].
  
\bibitem{Lee:2016cgz}
R.~N.~Lee, A.~V.~Smirnov, V.~A.~Smirnov and M.~Steinhauser,
Phys. Rev. D \textbf{94}, no.5, 054029 (2016)
[arXiv:1608.02603 [hep-ph]].

\bibitem{Kataev:2015yha}
A.~L.~Kataev and V.~S.~Molokoedov,
Phys. Rev. D \textbf{92}, no.5, 054008 (2015)
[arXiv:1507.03547 [hep-ph]].

\bibitem{Garkusha:2018mua} 
  A.~V.~Garkusha, A.~L.~Kataev and V.~S.~Molokoedov,
  JHEP {\bf 1802}, 161 (2018)
  [arXiv:1801.06231 [hep-ph]].

\bibitem{Crewther:1972kn} 
  R.~J.~Crewther,
  Phys.\ Rev.\ Lett.\  {\bf 28}, 1421 (1972).

\bibitem{Broadhurst:1993ru} 
  D.~J.~Broadhurst and A.~L.~Kataev,
  Phys.\ Lett.\ B {\bf 315}, 179 (1993)
  [hep-ph/9308274].

\bibitem{Baikov:2010je} 
  P.~A.~Baikov, K.~G.~Chetyrkin and J.~H.~Kuhn,
  Phys.\ Rev.\ Lett.\  {\bf 104}, 132004 (2010)
  [arXiv:1001.3606 [hep-ph]].

\bibitem{Kataev:2010du}
A.~L.~Kataev and S.~V.~Mikhailov,
Theor. Math. Phys. \textbf{170}, 139-150 (2012)
[arXiv:1011.5248 [hep-ph]].

\bibitem{Cvetic:2016rot}
G.~Cveti\v{c} and A.~L.~Kataev,
Phys. Rev. D \textbf{94}, no.1, 014006 (2016)
[arXiv:1604.00509 [hep-ph]].

\bibitem{Gabadadze:2017ujx}
G.~Gabadadze and G.~Tukhashvili,
Phys. Lett. B \textbf{782}, 202-209 (2018)
[arXiv:1712.09921 [hep-th]].

\bibitem{Baikov:2022zvq}
P.~A.~Baikov and S.~V.~Mikhailov,
JHEP \textbf{09}, 185 (2022)
[arXiv:2206.14063 [hep-ph]].

\bibitem{Chetyrkin:2022fqk}
K.~G.~Chetyrkin,
Nucl. Phys. B \textbf{985}, 115988 (2022)
[arXiv:2206.12948 [hep-ph]].

\bibitem{Gabadadze:1995ei} 
  G.~T.~Gabadadze and A.~L.~Kataev,
  JETP Lett.\  {\bf 61}, 448 (1995)
  [Pisma Zh.\ Eksp.\ Teor.\ Fiz.\  {\bf 61}, 439 (1995)]
  [hep-ph/9502384].

\bibitem{Crewther:1997ux} 
  R.~J.~Crewther,
  Phys.\ Lett.\ B {\bf 397}, 137 (1997)
  [hep-ph/9701321].

\bibitem{Braun:2003rp} 
  V.~M.~Braun, G.~P.~Korchemsky and D.~Müller,
  Prog.\ Part.\ Nucl.\ Phys.\  {\bf 51}, 311 (2003)
  [hep-ph/0306057].

\bibitem{Kataev:2017oqg}
A.~L.~Kataev and V.~S.~Molokoedov,
J. Phys. Conf. Ser. \textbf{938}, no.1, 012050 (2017)
[arXiv:1711.03997 [hep-ph]].

\bibitem{Molokoedov:2020kxq}
V.~S.~Molokoedov, PhD thesis, 2020. In Russian.
\url{https://www.inr.ru/rus/referat/molokoed/dis.pdf}

\bibitem{vonSmekal:2009ae} 
  L.~von Smekal, K.~Maltman and A.~Sternbeck,
  Phys.\ Lett.\ B {\bf 681}, 336 (2009)
  [arXiv:0903.1696 [hep-ph]].

\bibitem{Gracey:2013sca} 
  J.~A.~Gracey,
  J.\ Phys.\ A {\bf 46}, 225403 (2013)
  [arXiv:1304.5347 [hep-ph]].

\bibitem{Gracey:2014pba} 
  J.~A.~Gracey,
  Phys.\ Rev.\ D {\bf 90}, no. 9, 094026 (2014)
  [arXiv:1410.6715 [hep-ph]].

\bibitem{Ruijl:2017eht} 
  B.~Ruijl, T.~Ueda, J.~A.~M.~Vermaseren and A.~Vogt,
  JHEP {\bf 1706}, 040 (2017)
  [arXiv:1703.08532 [hep-ph]].

\bibitem{Zeng:2020lwi}
J.~Zeng, X.~G.~Wu, X.~C.~Zheng and J.~M.~Shen,
Chin. Phys. C \textbf{44}, no.11, 113102 (2020)
[arXiv:2004.12068 [hep-ph]].

\bibitem{Gracey:2022xjs}
J.~A.~Gracey and R.~H.~Mason,
[arXiv:2210.14604 [hep-ph]].

\bibitem{Grozin:2022umo}
A.~Grozin,
[arXiv:2212.05290 [hep-ph]].

\bibitem{Brodsky:1999fr} 
  S.~J.~Brodsky, M.~Melles and J.~Rathsman,
  Phys.\ Rev.\ D {\bf 60}, 096006 (1999)
  [hep-ph/9906324].

\bibitem{Brodsky:1994eh}
S.~J.~Brodsky and H.~J.~Lu,
Phys. Rev. D \textbf{51}, 3652-3668 (1995)
[arXiv:hep-ph/9405218 [hep-ph]].

\bibitem{Kiselev:2002iy}
V.~V.~Kiselev, A.~K.~Likhoded, O.~N.~Pakhomova and V.~A.~Saleev,
Phys. Rev. D \textbf{66}, 034030 (2002)
[arXiv:hep-ph/0206140 [hep-ph]].

\bibitem{Deur:2016tte} 
  A.~Deur, S.~J.~Brodsky and G.~F.~de Teramond,
  Prog.\ Part.\ Nucl.\ Phys.\  {\bf 90}, 1 (2016)
  [arXiv:1604.08082 [hep-ph]].

\bibitem{Hoque:2020qak}
R.~Hoque, B.~J.~Hazarika and D.~K.~Choudhury,
Eur. Phys. J. C \textbf{80}, no.12, 1213 (2020)

\bibitem{Afonin:2022aqt}
S.~Afonin and T.~Solomko,
J. Phys. G \textbf{49}, no.10, 105003 (2022)
[arXiv:2208.02604 [hep-ph]].

\bibitem{Grunberg:1982fw}
  G.~Grunberg,
  Phys.\ Rev.\ D {\bf 29} (1984) 2315.
  
\bibitem{Krasnikov:1981rp}
  N.~V.~Krasnikov,
  Nucl.\ Phys.\ B {\bf 192} (1981) 497.
  
\bibitem{Kataev:1981aw}
A.~L.~Kataev, N.~V.~Krasnikov and A.~A.~Pivovarov,
Phys. Lett. B \textbf{107}, 115-118 (1981)
  
\bibitem{Gross:1973id}
  D.~J.~Gross and F.~Wilczek,
  Phys.\ Rev.\ Lett.\  {\bf 30} (1973) 1343.
  
\bibitem{Politzer:1973fx}
  H.~D.~Politzer,
  Phys.\ Rev.\ Lett.\  {\bf 30} (1973) 1346.
  
\bibitem{Stevenson:1981vj} 
  P.~M.~Stevenson,
  Phys.\ Rev.\ D {\bf 23}, (1981) 2916.

\bibitem{Kataev:1995vh}
  A.~L.~Kataev and V.~V.~Starshenko,
  Mod.\ Phys.\ Lett.\ A {\bf 10} (1995) 235
  [hep-ph/9502348].

\bibitem{Jones:1974mm}
  D.~R.~T.~Jones,
  Nucl.\ Phys.\ B {\bf 75} (1974) 531.
  
\bibitem{Caswell:1974gg}
  W.~E.~Caswell,
  Phys.\ Rev.\ Lett.\  {\bf 33} (1974) 244.

\bibitem{Egorian:1978zx}
  E.~Egorian and O.~V.~Tarasov,
  Teor.\ Mat.\ Fiz.\  {\bf 41} (1979) 26
   [Theor.\ Math.\ Phys.\  {\bf 41} (1979) 863].

\bibitem{Tarasov:1980au}
  O.~V.~Tarasov, A.~A.~Vladimirov and A.~Y.~Zharkov,
  Phys.\ Lett.\ B {\bf 93} (1980) 429.

\bibitem{Larin:1993tp}
  S.~A.~Larin and J.~A.~M.~Vermaseren,
  Phys.\ Lett.\ B {\bf 303} (1993) 334
  [hep-ph/9302208].

\bibitem{vanRitbergen:1997va}
  T.~van Ritbergen, J.~A.~M.~Vermaseren and S.~A.~Larin,
  Phys.\ Lett.\ B {\bf 400} (1997) 379
  [hep-ph/9701390].

\bibitem{Czakon:2004bu}
  M.~Czakon,
  Nucl.\ Phys.\ B {\bf 710} (2005) 485
  [hep-ph/0411261].
  
\bibitem{Smirnov:2009fh}
A.~V.~Smirnov, V.~A.~Smirnov and M.~Steinhauser,
Phys. Rev. Lett. \textbf{104}, 112002 (2010)
[arXiv:0911.4742 [hep-ph]].
  
\bibitem{Anzai:2009tm}
C.~Anzai, Y.~Kiyo and Y.~Sumino,
Phys. Rev. Lett. \textbf{104}, 112003 (2010)
[arXiv:0911.4335 [hep-ph]].
  
\bibitem{PSLQ}
H.~R.~P.~Ferguson and D.~H.~Bailey, 
 RNR Technical Report RNR-91-032.
\url{https://www.nas.nasa.gov/assets/pdf/techreports/1991/rnr-91-032.pdf}

\bibitem{Bailey:1999nv} 
  D.~H.~Bailey and D.~J.~Broadhurst,
  Math.\ Comput.\  {\bf 70}, 1719 (2001)
  [math/9905048 [math-na]].

\bibitem{Lee:2009dh} 
  R.~N.~Lee,
  Nucl.\ Phys.\ B {\bf 830}, 474 (2010)
  [arXiv:0911.0252 [hep-ph]].

\bibitem{Lee:2015eva} 
  R.~N.~Lee and K.~T.~Mingulov,
  Comput.\ Phys.\ Commun.\  {\bf 203}, 255 (2016)
  [arXiv:1507.04256 [hep-ph]].

\bibitem{Nesterenko:2017wpb}
A.~V.~Nesterenko,
Eur. Phys. J. C \textbf{77}, no.12, 844 (2017)
[arXiv:1707.00668 [hep-ph]].

\bibitem{Davier:2023hhn}
M.~Davier, D.~D\'\i{}az-Calder\'on, B.~Malaescu, A.~Pich, A.~Rodr\'\i{}guez-S\'anchez and Z.~Zhang,
[arXiv:2302.01359 [hep-ph]].

\bibitem{Chetyrkin:1979bj} 
  K.~G.~Chetyrkin, A.~L.~Kataev and F.~V.~Tkachov,
Phys.\ Lett.\  {\bf 85B}, 277 (1979).

\bibitem{Dine:1979qh} 
  M.~Dine and J.~R.~Sapirstein,
  Phys.\ Rev.\ Lett.\  {\bf 43}, 668 (1979).

\bibitem{Celmaster:1979xr} 
  W.~Celmaster and R.~J.~Gonsalves,
Phys.\ Rev.\ Lett.\  {\bf 44}, 560 (1980).

\bibitem{Gorishnii:1990vf} 
  S.~G.~Gorishnii, A.~L.~Kataev and S.~A.~Larin,
Phys.\ Lett.\ B {\bf 259}, 144 (1991).

\bibitem{Surguladze:1990tg} 
L.~R.~Surguladze and M.~A.~Samuel,
  Phys.\ Rev.\ Lett.\  {\bf 66}, 560 (1991)
  Erratum: [Phys.\ Rev.\ Lett.\  {\bf 66}, 2416 (1991)].

\bibitem{Baikov:2008jh} 
  P.~A.~Baikov, K.~G.~Chetyrkin and J.~H.~Kuhn,
  Phys.\ Rev.\ Lett.\  {\bf 101}, 012002 (2008)
[arXiv:0801.1821 [hep-ph]].
  
\bibitem{Baikov:2012zn} 
  P.~A.~Baikov, K.~G.~Chetyrkin, J.~H.~Kuhn and J.~Rittinger,
  Phys.\ Lett.\ B {\bf 714}, 62 (2012)
  [arXiv:1206.1288 [hep-ph]].
 
\bibitem{Workman:2022ynf}
R.~L.~Workman \textit{et al.} [Particle Data Group],
PTEP \textbf{2022}, 083C01 (2022)

\bibitem{Kodaira:1978sh} 
  J.~Kodaira, S.~Matsuda, T.~Muta, K.~Sasaki and T.~Uematsu,
  Phys.\ Rev.\ D {\bf 20}, 627 (1979).
 
\bibitem{Gorishnii:1985xm}
  S.~G.~Gorishnii and S.~A.~Larin,
  Phys.\ Lett.\ B {\bf 172} (1986) 109.
 
\bibitem{Larin:1991tj}
  S.~A.~Larin and J.~A.~M.~Vermaseren,
  Phys.\ Lett.\ B {\bf 259} (1991) 345.
 
\bibitem{Larin:2013yba} 
  S.~A.~Larin,
  Phys.\ Lett.\ B {\bf 723}, 348 (2013)
  [arXiv:1303.4021 [hep-ph]].
 
\bibitem{Baikov:2015tea} 
  P.~A.~Baikov, K.~G.~Chetyrkin and J.~H.~Kühn,
  Nucl.\ Part.\ Phys.\ Proc.\  {\bf 261-262}, 3 (2015)
  [arXiv:1501.06739 [hep-ph]].
 
\bibitem{Beneke:1998ui} 
  M.~Beneke,
  Phys.\ Rept.\  {\bf 317}, 1 (1999)
  [hep-ph/9807443].
  
\bibitem{Zakharov:1992bx}
V.~I.~Zakharov,
Nucl. Phys. B \textbf{385}, 452-480 (1992)
 
\bibitem{Chetyrkin:2000dq} 
  K.~G.~Chetyrkin and A.~Retey,
  hep-ph/0007088.
  
\bibitem{Jegerlehner:2005fs}
F.~Jegerlehner and O.~V.~Tarasov,
Phys. Lett. B \textbf{639}, 299-306 (2006)
[arXiv:hep-ph/0510308 [hep-ph]].
  
\bibitem{'t Hooft 1}
G.~'t~Hooft, ``Some observations in quantum chromodynamics,'' Notes based on lectures given at Orbis Scientiae. January 17-21, 1977. University of Miami, Coral Gables, Floride (Reprint of February 1977).
  
\bibitem{'t Hooft 2}
G.~'t~Hooft, ``Can We Make Sense Out of Quantum Chromodynamics?,'' Lectures given at
Int. School of Subnuclear Physics, Erice, Sisily. July 23 - August 10, 1977. PRINT-77-0723
(UTRECHT) Subnucl. Ser. {\bf{15}} (1979) 943.
  
\bibitem{Garkusha:2011xb} 
  A.~V.~Garkusha and A.~L.~Kataev,
  Phys.\ Lett.\ B {\bf 705}, 400 (2011)
  [arXiv:1108.5909 [hep-ph]].
 
\bibitem{Gorishnii:1990kd}
  S.~G.~Gorishnii, A.~L.~Kataev, S.~A.~Larin and L.~R.~Surguladze,
  Phys.\ Lett.\ B {\bf 256} (1991) 81.
 
\bibitem{Baikov:2012zm}
  P.~A.~Baikov, K.~G.~Chetyrkin, J.~H.~Kuhn and J.~Rittinger,
  JHEP {\bf 1207} (2012) 017
  [arXiv:1206.1284 [hep-ph]].
  
\bibitem{Herzog:2017ohr}
F.~Herzog, B.~Ruijl, T.~Ueda, J.~A.~M.~Vermaseren and A.~Vogt,
JHEP \textbf{02}, 090 (2017)
[arXiv:1701.01404 [hep-ph]].

\bibitem{Luthe:2017ttg}
T.~Luthe, A.~Maier, P.~Marquard and Y.~Schroder,
JHEP \textbf{10}, 166 (2017)
[arXiv:1709.07718 [hep-ph]].

\bibitem{Kataev:2013vua}
A.~L.~Kataev,
JHEP \textbf{02}, 092 (2014)
[arXiv:1305.4605 [hep-th]].
 
\bibitem{Blumlein:2009cf} 
  J.~Blumlein, D.~J.~Broadhurst and J.~A.~M.~Vermaseren,
  Comput.\ Phys.\ Commun.\  {\bf 181}, 582 (2010)
  [arXiv:0907.2557 [math-ph]].
 
\bibitem{Blumlein:2009fz}
J.~Bl\"umlein,
Clay Math. Proc. \textbf{12}, 167-188 (2010)
[arXiv:0901.0837 [math-ph]].

\bibitem{Anzai:2012xw} 
  C.~Anzai and Y.~Sumino,
  J.\ Math.\ Phys.\  {\bf 54}, 033514 (2013)

\bibitem{Kompaniets:2017yct}
M.~V.~Kompaniets and E.~Panzer,
Phys. Rev. D \textbf{96}, no.3, 036016 (2017)
[arXiv:1705.06483 [hep-th]].


\bibitem{Michael:1986yi} 
  C.~Michael,
  Nucl.\ Phys.\ B {\bf 280}, 13 (1987).

\bibitem{Rothe:1995hu} 
  H.~J.~Rothe,
  Phys.\ Lett.\ B {\bf 355}, 260 (1995)
  [hep-lat/9504012].

\bibitem{Dosch:1995fz} 
  H.~G.~Dosch, O.~Nachtmann and M.~Rueter,
  hep-ph/9503386.

\bibitem{Shoshi:2002rd} 
  A.~I.~Shoshi, F.~D.~Steffen, H.~G.~Dosch and H.~J.~Pirner,
  Phys.\ Rev.\ D {\bf 68}, 074004 (2003)
  [hep-ph/0211287].
  
\bibitem{Chernodub:2010sq}
M.~N.~Chernodub,
Universe \textbf{6}, no.11, 202 (2020)
[arXiv:1003.3225 [hep-ph]].

\bibitem{Adler:1976zt} 
  S.~L.~Adler, J.~C.~Collins and A.~Duncan,
  Phys.\ Rev.\ D {\bf 15}, 1712 (1977).

\bibitem{Collins:1976yq} 
  J.~C.~Collins, A.~Duncan and S.~D.~Joglekar,
  Phys.\ Rev.\ D {\bf 16}, 438 (1977).

\bibitem{Nielsen:1977sy} 
  N.~K.~Nielsen,
  Nucl.\ Phys.\ B {\bf 120}, 212 (1977).

\end{thebibliography}
\end{document}